\documentclass[12pt]{article}
\usepackage[utf8]{inputenc}
\usepackage[english]{babel}

\usepackage{graphicx}% Include figure files
\usepackage{bm}% bold math
\usepackage{latexsym}
\usepackage{dcolumn}

\usepackage{tensor}

\usepackage{amsmath}
\usepackage{epsfig,amssymb,euscript,mathrsfs}
\usepackage{array,calc,epsfig}

\usepackage{amsfonts}
\usepackage{amsthm}
\usepackage{slashed}
\usepackage{caption}
\usepackage{pdfpages}
\usepackage{braket}
\usepackage{rotating}
\usepackage{bbm}

\usepackage{tikz-cd}

\usepackage[noadjust]{cite}
\usepackage{chngcntr}
\usepackage[pdfpagelabels]{hyperref}
\usepackage{color}

\usepackage{booktabs}
\usepackage{comment}
\usepackage{cases} %per scrivere sistemi di equazioni labellate
\usepackage{empheq}
\usepackage{cancel}

\usepackage{xfrac} %per frazioni storte (\sfrac{}{})
\usepackage{float} %per \begin{figure}[H]
\usepackage{subfig}

\def\ra{\rightarrow}
\def\lra{\leftrightarrow}
\newcommand{\numberset}{\mathbb}

\newcommand{\R}{\numberset{R}}

\newcommand{\Z}{\numberset{Z}}

\newcommand{\p}{\partial}

\newcommand{\T}{\mathcal{T}}

\newcommand{\A}{\mathcal{A}}

\renewcommand{\d}{\mathrm{d}}

\renewcommand{\Im}{\text{Im}}

\newcommand{\spin}{\text{Spin}}
\newcommand{\SO}{\text{SO}}
\newcommand{\U}{\text{U}}
\newcommand{\SL}{\text{SL}}
\newcommand{\SU}{\text{SU}}

\newcommand{\arf}{\text{Arf}}

\newcommand{\Tr}{\text{Tr}}

\def\be{\begin{equation}} 
\def\ee{\end{equation}}

\def\a{\alpha} 
\def\b{\beta} 
\def\g{\gamma} 
\def\G{\Gamma} 
\def\D{\Delta}

\def\z{\zeta}

\def\l{\lambda}

\def\s{\sigma} 
\def\S{\Sigma} 
\def\t{\tau} 
\def\f{\phi} 
\def\vf{\varphi}

\def\w{\omega} 
 
\def\Th{\Theta} 
\def\th{\theta}
\def\wt{\widetilde} 
\def\wh{\widehat}

\numberwithin{equation}{section}

\newlength\dlf  % Define a new measure, dlf

\begin{document}

\begin{titlepage} 
\begin{center} 
\vspace{1cm}
{\LARGE Bosonizations and dualities in 2+1 dimensions}
\vspace{1cm}

Andrea Cappelli${}^a$ and Riccardo Villa${}^{a,b}$

\vspace{1cm}
{\em ${}^{a}$INFN, Sezione di Firenze,\\
  Via G. Sansone 1, 50019 Sesto Fiorentino - Firenze, Italy}\\

\bigskip
{\em ${}^{b}$Dipartimento di Fisica, Universit\`a di Firenze,\\
Via G. Sansone 1, 50019 Sesto Fiorentino - Firenze, Italy} \\
\end{center} 

\vspace{1cm}

\begin{abstract}
 We discuss two methods for relating bosonic and fermionic
  relativistic field theories in 2+1 dimensions, the $\Z_2^f$ gauging
  and the flux attachment. The first is primarily a correspondence
  between topological theories. It amounts to summing over fermionic
  spin structures, as is familiar in two-dimensional conformal
  theories. Its inverse map, fermionization, shows how spin structures
  and $\Z_2^f$ fermion parity emerge from a bosonic theory equipped
  with a dual $\Z_2^{(1)}$ generalized symmetry.  The second method,
  flux attachment, gives spin and statistics to charged particles by
  coupling them to a Chern-Simons theory, and provides the basis for
  the Abelian dualities.  We illustrate the two bosonizations with
  explicit results in a solvable semiclassical conformal theory, and show
  their differences and interplays with particle-vortex dualities.  We
  employ the so-called loop model, which can
  describe general infrared critical points in 2+1 dimensions in
  the semiclassical limit.  We also combine the two bosonizations
  to obtain further duality relations. By applying $\Z_2^f$ gauging to
  the Dirac-boson and Majorana-boson flux-attachment dualities, we
  find new relations between bosonic theories.
\end{abstract} 
 
\vfill 
\end{titlepage} 
\pagenumbering{arabic}

\tableofcontents
\newpage

%-1----------------------------------------------

\section{Introduction}

Bosonization, or its inverse, fermionization, is a well-known exact map
between two 1+1 dimensional field theories,  involving
bosonic and fermionic  degrees of freedom, respectively.
This correspondence is very useful because it entails a number of
non-perturbative results.
In presence of conformal invariance, it amounts to the explicit relabeling
of free-particle states in the Hilbert space, originating from
the Jacobi triple-product identity \cite{ginspargCFT}. 
Let us recall two aspects of this map:
\begin{itemize}
\item
  The relation between Hilbert spaces, which can be summarized by
  the partition functions on the 1+1d torus, 
  \be
Z_b=\sum_\eta Z_f[\eta]=Z_{NS}+Z_{\wt{NS}}+Z_R+Z_{\wt{R}}.
  \label{Z-sum}
  \ee
Here, the bosonic partition function $Z_b$ is obtained by summing the
  fermionic ones $Z_f[\eta]$ over the spin structures
  $\eta=NS,\wt{NS},R, \wt{R} $ of the torus. In Hamiltonian
  formulation, this amounts to the trace $\Tr[(1+(-1)^F)e^{-\b H}]$, i.e. to
  `gauging' the $\Z_2^f$ fermion parity symmetry. Clearly, the two
  distinctive fermionic features, of spin structures and parity symmetry, are
  washed out in the bosonic theory.
\item 
The possibility to express the fermionic field in bosonic form, 
\be
\psi(x) = \exp(i\vf(x)),
\label{f-field}
\ee
where $\psi$ is the Dirac fermion and  $\vf$ the compactified boson.
In particular, the fermion field is characterized
by non-trivial phase factors for spin and statistics.
In this respect, bosonization is a particular kind of a duality, since
it allows to express the path-integral in terms of two different
sets of variables.
\end{itemize}

In this work, we discuss the present understanding of bosonization
in 2+1 dimensional relativistic field theory.
Of course, we should not expect exact maps between free-particle Fock states
and fields. Nonetheless, explicit results will be obtained at the
topological level, and also in presence of a relatively
simple, semiclassical dynamics.

It turns out that the two aspects of 1+1d bosonization just described
become distinct in higher dimensions, leading to independent
maps, which will be referred to as
$\Z^f_2$ gauging and flux attachment.

\vspace{0.5cm}

\noindent {\bf $\Z^f_2$ gauging}

\noindent The relation \eqref{Z-sum} between partition functions has been generalized
at the topological level in any spacetime dimensions $d$, in
particular for $d=3,4$: it involves a `spin-TQFT', i.e. fermionic
topological field theory, and its bosonic `shadow theory'
\cite{gaiottokapustinspinTQFT1,gaiottokapustinspinTQFT2}. The map
is invertible, between fermionic spin sectors $Z_f[\eta]$ and bosonic
`twisted sectors' $Z_b[B]$, in presence of a $(d-1)$-form gauge field
$B$ taking values in $\Z_2$.  While the sum over spin structures
\eqref{Z-sum} amounts to gauging the $\Z^f_2$ fermion parity, the inverse
map corresponds to gauging the dual $\Z_2^{(d-2)}$ generalized
symmetry of the bosonic theory.  This bosonization method will be
equivalently called ``$\Z^f_2$ gauging'' or ``sum over spin structures''.

In this work, we discuss the topological map in rather simple terms,
and also extend it in the presence of a solvable dynamics.  We rely on
our earlier analysis of the loop model \cite{3dbosonization2024}, a
2+1d conformal theory first formulated for massless excitations at the
boundary of 3+1d topological insulators \cite{loopmodel}.  This theory
can describe the infrared critical point of most 2+1d theories, such
as QED$_3$ and QED$_{4,3}$ \cite{loopmodel}, within the semiclassical
approximation. It possesses a critical line and solitonic states with
electric and magnetic charges, and its partition function can be
explicitly obtained on the cylinder $\R\times S^2$, giving access to
the conformal spectrum.  With the help of results in this theory, we
illustrate the $\Z^f_2$ gauging \eqref{Z-sum} in 2+1 dimensions and
the relation it entails between bosonic and fermionic partition
functions.

\vspace{0.5cm}

\noindent {\bf Flux attachment}

\noindent The flux attachment is a well-known method for changing the statistics
of electrons in 2+1 dimensional nonrelativistic many-body
systems. It is realized by coupling the particles
to a gauge field with Chern-Simons action: this leads to a two-body potential,
which can be eliminated by an instantaneous gauge transformation
of the wave function $\Psi$, as follows
\begin{align}
  & \Psi(z_1,\cdots, z_N) \ \to\
    \exp\left(i\sum_{1 \le i<j\le N} \Th (z_i,z_j) \right)
    \Psi(z_1,\cdots, z_N),
\nonumber  \\
&  \Th(z_i,z_j) = \Im\log(z_i-z_j),
\label{flux-NR}
\end{align}
where $z_i$, $i=1,\dots,N$ are the particle positions in complex
coordinates.  The phase factor $\Th(z_i,z_j)$ changes the statistics
of particles from bosonic to fermionic and viceversa.  This
correspondence is much used and well understood in the physics e.g. of
the quantum Hall effect \cite{cappellianomaliescondmat}, where it
received experimental confirmation.  Although singular, the
wavefunction transformation \eqref{flux-NR} is well-defined in
presence of a gap for excitations.

In relativistic theories, coupling bosonic degrees of freedom to the Chern-Simons theory
also introduces Aharonov-Bohm phases, as well as the dependence on
spin structure and fermion parity, which are actually needed for the
definition of this theory \cite{Witten2003SL2Z,wittenwebofduality}.
In this sense, the flux attachment does map a bosonic theory into a
fermionic one. However, the quantum corrections are more relevant, and
it is not always known how to rewrite the resulting theory in terms of
(interacting) fermion fields, as in \eqref{f-field}, or whether this
is possible at all.

A case in which the fermion description can be found
after flux attachment is the following
\begin{equation}
\label{dirac-fa}
{\cal L}_f=\Bar{\psi} i \slashed{D}_A \psi
\qquad \longleftrightarrow \qquad
{\cal L}_b=
|D_a\phi|^2 + V(|\phi|)+ \frac{i}{4\pi} a\d a + \frac{i}{2\pi} a \d A\,.
\end{equation} 
In this expression, on the left-hand side a free fermion is coupled to
the electromagnetic background A, while on
the right-hand side a charged scalar is self-interacting
with potential $V(|\f|)$ and coupled the the Chern-Simons field $a$,
which itself is
electric-magnetic dual with respect to the $A$ background.
As better described in Section \ref{fluxattachmentSEC},
this correspondence is meant to
hold in the infrared limit, when the scalar self-interaction is tuned
to the Wilson-Fisher critical point, given that the fermion is massless.
This infrared duality has been conjectured and verified numerically
\cite{3ddualityreviewSenthilSon}. It also gives the seed for other dualities
\cite{wittenwebofduality,tongwebofduality} as it
will be recalled later.

\vspace{0.5cm}

\noindent Given this state of the art, it is interesting to investigate
the flux attachment in the loop model, which allows for exact calculations,
and compare the results obtained by the other method of $\Z_2^f$ gauging.
This program is carried out in the present work.

The outcome is that the two bosonization procedures give different results
for the effective response actions, the conformal spectra and
the partition functions. Nonetheless, both resulting fermionic theories
obey particle-vortex duality relations,
inherited from the self-duality of the original bosonic loop model.
All boson-fermion and particle-vortex dualities are generalized
to the case of excitations with fractional charge, by allowing
Chern-Simons couplings $k\neq 1$ in \eqref{dirac-fa}.

With hindsight, there is no reason to expect that the two fermionization
methods in 2+1 dimensions
should give equal results. They both introduce a dependence on spin
structures that the bosonic theory does not have.
The $\Z^f_2$ discrete gauging
is topological and only causes a parity rule in the original
spectrum of electric and magnetic charges.
Instead, flux attachment, i.e. Chern-Simons $\U(1)$ gauging, adds
dynamics to the loop model and constraints the spectrum more strongly,
by forcing electric and magnetic charges to be equal.

We remark that the two bosonization/fermionization maps can only be
realized under some conditions.  Given a fermionic theory, it is
always possible to obtain the bosonic counterpart by summing over spin
structures. For the inverse map to exist, the bosonic theory should
have the $\Z_2^{(1)}$ generalized symmetry, dual of $\Z^f_2$, with a
specific form of $\Z_2$ anomaly \cite{gaiottokapustinspinTQFT1}.  In
the flux attachment \eqref{dirac-fa}, the bosons should be charged.
Nonetheless, we shall also discuss the bosonization of Majorana
fermions
\cite{metlitski2017MajoranaDuality,AharonyBeniniHsinSeiberg2017CSSODualities}.

Being different maps, the flux attachment \eqref{dirac-fa} and $\Z_2^f$
gauging can be combined together for obtaining other dualities.
Starting from the Dirac-scalar duality \eqref{dirac-fa}, the $\Z_2^f$
gauging of both sides yields a new boson-boson correspondence.
The same analysis is carried on the Majorana-boson duality.

The outline of this work is as follows.  In Section
\ref{bosonizationSEC}, we introduce the first bosonization method,
based on $\Z^f_2$ gauging, starting from the analysis of topological
theories in $d$ dimensions, and then more specifically in $2+1d$.  In
Section \ref{fluxattachmentSEC}, we review the flux attachment method
and the associated particle-vortex dualities. In Section \ref{loop model sec}, we recall the features of the loop
model, its bosonic spectrum and selfduality. We then summarize the
fermionization by $\Z_2^{(1)}$ gauging obtained in the earlier work
\cite{3dbosonization2024}, leading to the fermionic partition
functions in the two spin sectors of the conformal cylinder
$\R\times S^2$.  We also describe the generalized vertex operators
which express the fermion field and its time reversal transformations,
extending the 1+1 dimensional relation \eqref{Z-sum}.  We then perform
the flux attachment on the same theory, find the second fermionic
theory and its partition function, and check its particle-vortex
duality. In Section \ref{Bosonization of FA sec}, we combine the two
bosonization maps and find new bosonic dualities. In the
Conclusions, we comment on future developments of bosonization in
3+1 dimensions.  

There are also two appendices. Appendix \ref{appBosonization} contains more
technical aspects of $\Z_2^f$ gauging, its comparison with the existing literature and its extension for twisted
spin structures and non-orientable manifolds. We summarize and unify in the same perspective results of Refs. \cite{gaiottokapustinspinTQFT1,gaiottokapustinspinTQFT2,bhardwaj2017unorientedTFT3d,thorngren2020anomalies,thorngren2018thesisPHD,tongArf,HsinShao2020,seibergwitten2016gappedTI,wittenwebofduality,backfiringbosonization2024}. Appendix \ref{app U4k to Uk} reports in detail the relation between bosonic and fermionic $\U(1)_k$ Chern-Simons theories, useful for the new duality in Section \ref{Bosonization of FA sec}.

%-2---------------------------------------------------------

\section{Bosonization by $\Z_2^f$ gauging}
\label{bosonizationSEC}

In this Section we recall the bosonization procedure introduced in
\cite{gaiottokapustinspinTQFT1, gaiottokapustinspinTQFT2}.  A review
of this approach was already given in our precedent work
\cite{3dbosonization2024}, but here we will be more explicit following
\cite{thorngren2020anomalies,thorngren2018thesisPHD}. Technical details
and extensions are deferred to Appendix \ref{appBosonization}.

\subsection{General dimension}

In this Section we will use differential forms with values in $\Z_n$,
which are more
properly described in terms of cochains in cohomology theory
\cite{HatcherAT,nakahara,kapustinseiberg2014}.
We assume that the continuum spacetime manifold $X$ has associated
a triangulation
with same topology, on which the simplicial calculus is realized.
This involves the $p$-cochains $\a_p \in C^p(X,\Z_n)$, discrete analog
of $p$-forms, which take values $0,1,...,n-1$ in
the additive group $\Z_n$. The simplicial and differential calculi go in
parallel: the coboundary operator $\delta$ corresponds to the exterior
derivative $d$, such that $\delta^2=0$, and the cup product $\cup$ is the
counterpart of the wedge product $\wedge$ on forms.

We shall equivalently use the word form for cochains and keep
the notation $\d$, $\d^2=0$ also for coboundary. Regarding the properties of
$\cup$, they are the same as those of $\wedge$ on elements of the
cohomology groups, which are the cohomology classes $[\a_p] \in H^p(X,\Z_n)$,
obeying $[\a_p]=0$ for $\a_p=\d\a_{p-1}$, in particular.
On single representatives of cohomology classes or non-closed
cochains $\a_p$, the $\cup$ product fails to be graded-commutative as $\wedge$,
by terms involving the higher cup product $\cup_1$
\cite{kapustinseiberg2014,gaiottokapustinspinTQFT1}.
This difference will not be important in the following discussion and
we will be pointed out when it occurs.

%-2.1.1------------------------------------

\subsubsection{Spin structures and fermion parity}

We start by recalling that fermions can be defined on spin manifolds
$X$, characterized by the vanishing of the second Stiefel--Whitney
class $[w_2(TX)] = [w_2]=0$, where $[w_2] \in H^2(X, \Z_2)$, and $TX$ is
the tangent bundle
\cite{nakahara,kapustinFermionicBordism}.  Actually, a non-vanishing
$[w_2]$ would provide an obstruction to lifting the SO($d$) frame
bundle to a Spin$(d)$ bundle.  Concretely, fermions can be globally
defined on the manifold when a condition for matching $\pm 1$ signs of
local quantities is fulfilled at triple intersections of patches
\cite{witten2016fermion}. When $[w_2] =0$ (i.e. $w_2$ is exact),
these signs can be chosen consistently. This amounts to a choice
of a spin structure on $X$. It follows that there is a globally
defined one-form $\eta$ valued in $\Z_2$ such that $w_2 = \d \eta$: we
can think of $\eta$ as the choice of the spin structure itself
\cite{thorngren2014framed, gaiottokapustinspinTQFT1,
  gaiottokapustinspinTQFT2, thorngren2018thesisPHD}.  Other
representatives $w_2=\d\l$ of the class $[w_2]=0$ correspond to
different local sign choices for the same spin bundle.

On a spin manifold $X$ there are generally several spin
structures. Indeed,  it is known that
given a reference spin structure $\eta$, we can find the other ones
by adding elements\footnote{Note the space of inequivalent spin
  structures on a manifold $X$ is an affine space over
  $H^1(X,\Z_2)$. The difference of any two spin structures is a $\Z_2$
  gauge field, $\eta-\eta' \in H^1(X, Z_2)$.}
$s\in H^1(X,\Z_2)$: $\eta'=\eta+s$.
In physical terms, different
spin structures amount to the choice of periodic or
antiperiodic boundary conditions for the fermion on the non-trivial
loops in $X$.
This is exactly the information contained in
$s \in H^1(X,\Z_2) \cong \text{Hom}(H_1(X),\Z_2)$
\cite{HatcherAT}. For every loop $\gamma$ in the homology of $X$, the
1-form $s$, obeying $\d s=0$, assigns the sign given by
\begin{equation}
\label{eta spin structure}
    (-1)^{\int_\gamma s} = \{-1,1\}, \qquad \int_\g s = \{0,1\}.
\end{equation}
We can associate the
$+1$ value to the natural choice of antiperiodic boundary conditions
and $-1$ to the periodic ones.
Note also that if $s = \d \chi$, the values of the
spin structure \eqref{eta spin structure} are not modified, since
$\int_\gamma \d \chi = 0$ on closed loops and on open lines it is just a
function of the endpoints: they are `gauge transformations', i.e. signs
that can be removed by local redefinitions of the fermion.
We recover the known fact that the number of inequivalent spin structures on
$X$ is given by the dimension of $H^1(X, \Z_2)$. 

The quantity $\eta+s$, with $s\in H^1(X,\Z_2)$, can be thought of as
being a 1-form gauge field with values in $\Z_2$, which can be
associated to a global zero-form $\Z_2$ symmetry.  Namely, a theory
that requires the choice of a spin structure $\eta$ is naturally
equipped with a $\Z_2$ symmetry. This should involve only fermions and
not bosons, which do not see $\eta$. An obvious candidate for this
symmetry is therefore the fermion parity $\Z_2^f$ generated by
$(-1)^F$.  It is then natural to interpret $\eta + s$ as gauge fields
for $\Z_2^f$. Note that another possibility is to consider $s$ alone
as the gauge field for $\Z_2^f$ (the standard choice for internal
symmetries \cite{dijkgraafwitten}). This different choice
is discussed in Appendix \ref{app gauge Z_2^f or spin str}.

%-2.1.2----------------------------
\subsubsection{Bosonization} 

We now turn to describe the bosonization approach
 in any spacetime dimensions $d$ \cite{thorngren2020anomalies}. Recall that
a fermionic theory depends on the choice of spin structure
of a given manifold $X$, while a bosonic theory depends only on its
orientation. Ignoring for simplicity any other dependence for now, we
denote the fermionic partition function as $Z_f[\eta]$. A
straightforward way to get a bosonic theory is to sum over all
the spin structures of $X$:
\begin{equation}
\label{oldbosonization}
    Z_b = \sum_\eta Z_f[\eta] = \sum_{s \in H^1(X,\Z_2)} Z_f[\eta_0+s],
\end{equation}
where in the last expression we fixed a reference (arbitrary) spin
structure $\eta_0$. This is the known
bosonization in 1+1 dimensions mentioned in the Introduction.
As described before, we
consider $\eta$ as the gauge field for fermion parity: therefore,
\eqref{oldbosonization} is really equivalent to gauging the fermion
parity symmetry by summing over its $\Z_2$ gauge field $\eta$. We call
this procedure $\Z_2^f$ gauging.

As a result, the bosonic theory
should have a dual $\Z_2^{(d-2)}$ symmetry\footnote{We use the
  superscript in $G^{(p)}$ to denote a
  $p$-form symmetry with group $G$ \cite{gaiotto2015generalized}.}
\cite{bhardwajTachikawa2018,3dbosonization2024}, generated by the Wilson lines of $\eta$. This symmetry can be probed
with a background gauge field
$B\in H^{d-1}(X,\Z_2)$. Turning
this field on, the sum \eqref{oldbosonization} becomes
\begin{equation}
\label{invertible bosonization}
Z_b[B] = \sum_\eta Z_f[\eta] (-1)^{\int_X \eta \cup B} =
(-1)^{\int_X \eta_0 \cup B} \sum_{s\in H^1(X,\Z_2)}
Z_f[\eta_0+s] (-1)^{\int_X s \cup B}.
\end{equation}
As usual for discrete symmetries, this map, being the analog of a Fourier transform, can be inverted by gauging the dual symmetry in the bosonic theory,
\begin{equation}
\label{fermionization}
        Z_f[\eta] = \sum_{B\in H^{d-1}(X,\Z_2)} Z_b[B] (-1)^{\int_X \eta \cup B},
\end{equation}
which is verified using the orthogonality
\begin{equation}
  \sum_{B\in H^{d-1}(X,\Z_2)} (-1)^{\int (\eta-\eta')\cup B} =
  \delta_{\eta\eta'\ {\rm mod}\ 2}.
\end{equation}
Summarizing, the $\Z^f_2$ gauging map is defined by the transformations
\eqref{invertible bosonization} and \eqref{fermionization}.

Everything said until now is pretty standard for gauging a
discrete symmetry. However, there is an important
observation to be made. In the bosonic theory \eqref{invertible
  bosonization}, the current $*\eta$ 
of the dual $\Z_2^{(d-2)}$ symmetry is not conserved in general,
because $\d \eta = w_2$: this indicates that the bosonic dual symmetry is
actually anomalous.

Indeed, by doing a background gauge transformation
$B \to B+\d \lambda$ in \eqref{invertible bosonization}, $Z_b[B]$ changes
by a phase:
\begin{equation}
\label{Z_b anomalous}
Z_b[B+\d \lambda] = Z_b[B] (-1)^{\int_X \eta \cup \d \lambda} =
Z_b[B] (-1)^{\int_X w_2 \cup \lambda} \,.
\end{equation}
In this expression, we used the Leibniz rule for $\d$, i.e.
$\d( \eta\cup\l)=\d \eta \cup \l - \eta \cup \d\l$, and 
integrated by parts assuming $X$ closed.

The non-invariance in \eqref{Z_b anomalous} is indeed
canceled by the introduction of the coupling $(-1)^{\int \eta \cup B}$ in
\eqref{fermionization}, which is therefore needed to gauge the bosonic
$\Z_2^{(d-2)}$ symmetry and recover $Z_f[\eta]$. This is expected, since
gauging $B$ without $\eta$ would give a fermionic theory without
background for $\Z_2^f$, i.e. $Z_f[\eta=0]$, but there is no zero in
the space of the spin structures. This implies that it is not possible
to gauge $B$ without introducing $\eta$.

The term $i\pi\int_X \eta\cup B$ canceling the anomaly in
\eqref{fermionization} is a local expression involving $\eta$,
i.e. the spin structure.  From the bosonic perspective, we have
introduced a peculiar extra field ($\eta$) to remove the anomalous
variation, transforming it in a fermionic theory. We thus say that
\eqref{Z_b anomalous} is an 't Hooft anomaly for the bosonic
$\Z_2^{(d-2)}$ symmetry and that this anomaly is trivialized by the
introduction of the spin structure $\eta$.

It is possible to give a standard description of the non-invariance
\eqref{Z_b anomalous} as an anomaly which obeys the inflow relation
from one dimension higher.
Assume that our spacetime $X$ is a boundary of the $d+1$ manifold $Y$,
$X =\p Y$, on which $B$ can be extended.
Then, we write the following topological action in $d+1$ dimensions 
\begin{equation}
\label{Z_2^(d-2) anomaly}
    \mathcal{A} = i \pi \int_Y w_2 \cup B = i \pi \int_Y Sq^2B \qquad (\text{mod $2\pi i$}).
  \end{equation}

Let us consider the first expression for $\mathcal{A}$ in
  \eqref{Z_2^(d-2) anomaly}.
When $Y$ is closed, this TQFT is correctly gauge
invariant under $B \ra B + \d\lambda$, owing to $\d w_2=0$;
when $Y$ has a boundary, the action is not invariant
 and its variation cancels the anomaly \eqref{Z_b anomalous} on $X=\p Y$.
This is the standard anomaly inflow relation.

The second expression for $\mathcal{A}$ in \eqref{Z_2^(d-2) anomaly}
follows from the identity
called Wu formula \cite{kapustinseiberg2014,kapustinthorngren2017}, 
and it involves the second Steenrod square $Sq^2$ of
the $B$ field.  The $p$-th Steenrod squares are homomorphisms between
the cohomology groups of $X$, $Sq^p: H^k(X,\Z_2)\to
H^{k+p}(X,\Z_2)$.  In particular, $Sq^2 B$ is quadratic in $B$.
These technical aspects are reviewed 
Appendix \ref{appBosonization}, together with the 
detailed comparison with the literature 
\cite{gaiottokapustinspinTQFT1,kapustinthorngren2017}.

The last form of the anomaly \eqref{Z_2^(d-2) anomaly} is convenient from
the bosonic perspective, since it does not explicitly involve  $\eta$.
Its expression cannot be written as a local
term in $d$ dimensions, thus making explicit the anomalous nature of the
$\Z_2^{(d-2)}$ symmetry that emerges in bosonic theories
which can be mapped into fermions. Thus, the action $i\pi\int_Y Sq^2 B$
defines a bosonic $\Z_2^{(d-2)}$ symmetry protected topological phase (SPT).
The first form for  $\mathcal{A}$ instead shows that this symmetry is no longer
anomalous when the $\eta$ field is introduced in the bosonic theory,
since $w_2=\d \eta$ on $X$.\footnote{
Further aspects of the two forms of the anomaly
\eqref{Z_2^(d-2) anomaly} are discussed in
Appendix \ref{appBosonization}.} 

The conclusion of this discussion is the following. Summing over the
spin structures in a fermionic theory is equivalent to gauging
$\Z_2^f$. This yields a bosonic theory with an
anomalous $\Z_2^{(d-2)}$ symmetry. This 't Hooft anomaly has a specific
form given by \eqref{Z_2^(d-2) anomaly} in any dimension $d$.
Conversely, a bosonic theory can mapped into a fermionic one, inverting
the sum over spin structures, if it
has a $\Z_2^{(d-2)}$ symmetry with anomaly \eqref{Z_2^(d-2)
  anomaly}. In this case, the dual fermionic theory is obtained by
gauging this symmetry after removing the anomaly.

%-2.1.3------------------------------
\subsubsection{Anyon condensation}

We conclude this Section by recalling a
useful physical picture to understand the relation between the fermionic and
bosonic theories and the anomaly \eqref{Z_2^(d-2) anomaly}
\cite{gaiottokapustinspinTQFT1,kapustinthorngren2017,3dbosonization2024}.

A $(d-2)$--form symmetry in $d$ dimensions is generated by line
operators, i.e. worldlines of (topological) quasiparticles. Gauging
this higher symmetry is equivalent to sum over the insertion
of such topological line defects in the correlation functions, which
means proliferating quasiparticle lines. This procedure is known
as ``condensation''. Actually, we can create condensates (states
with constant wave function) of bosonic particles, since other
statistics are not compatible with it. So, a $(d-2)$--form symmetry in
$d$ dimensions can be gauged if it is generated by bosonic particles,
otherwise is anomalous.

This is what happens for the $\Z_2^{(d-2)}$ symmetry \eqref{Z_b anomalous}.
Its anomaly \eqref{Z_2^(d-2) anomaly} signals that
this symmetry is generated by fermionic quasiparticles which cannot
condensate. To gauge this symmetry, we need to introduce the spin
structure $\eta$, which contributes with neutral fermionic probes as
\begin{equation}
    e^{\oint_\g \eta}.
\end{equation}
These neutral fermionic lines dress the fermionic generators of
$\Z_2^{(d-2)}$, creating bosonic pairs. These dressed lines can then
condense (equivalently, in the partition function, this is the
statement that \eqref{fermionization} can be gauged, since the anomaly
is removed by the coupling with $\eta$).

From this picture we learn that the anomaly \eqref{Z_2^(d-2) anomaly}
implies that in the bosonic theory there should be a quasiparticle
with fermionic self--statistic which fuses with itself into the
identity ($\Z_2^{(d-2)}$ symmetry).  After gauging, these fermionic
lines become transparent\footnote{In the sense that they act trivially
  on any observable, by definition of gauging.} and liberate
endpoints, which are the local fermions \cite{gaiottokapustinspinTQFT1}.
The universality of the anomaly
\eqref{Z_2^(d-2) anomaly}, i.e. its independence from
the specific theory considered, also follows
\cite{gaiottokapustinspinTQFT1}. Consider two different
decoupled theories QFT1 and QFT2 with fermionic quasiparticles
$\psi_1$ and $\psi_2$, then the product $\psi_1\psi_2$ is a bosonic
particle which generates a non anomalous $\Z_2^{(d-2)}$ symmetry; thus
the two anomalies are the same (two equal signs that square to one).

In two--dimensions, where there is no real difference between fermions
and bosons, fermions can also condense and, therefore, we expect that
there is no anomaly for the dual bosonic (zero-form) $\Z_2$
symmetry. This is indeed the case, since $Sq^2 B = 0$ if $B$ is a
one-form and thus the anomaly \eqref{Z_2^(d-2) anomaly} vanishes. This
does not mean that \eqref{fermionization} does not apply in $d=2$: to
recover the fermionic theory one should always gauge the dual $\Z_2$
symmetry with the spin structure insertion \eqref{fermionization}
\cite{thorngren2020anomalies,tongArf}.
The two-dimensional case is discussed in Section
\ref{FA vs Z2 gauging in 2d} and Appendix \ref{app gauge Z_2^f or spin str} in more detail.

%-2.2--------------------------------------------------

\subsection{The $2+1$ dimensional case}
We now specialize to  three
dimensions, where every orientable manifold is also a spin manifold.
The dual bosonic symmetry is an one-form symmetry $\Z_2^{(1)}$ and its
anomaly \eqref{Z_2^(d-2) anomaly} has the following explicit form
in terms of the 2-form field $B$,\footnote{In the following,
  equations involving $\Z_2$-valued forms are valid mod 2.}
\begin{equation}
\label{3d Sq2B}
    \mathcal{A} = i\pi \int_Y Sq^2 B = i\pi \int_Y B\cup B \,,
  \end{equation}
  where $Y$ is four-dimensional, $\p Y = X$.
  The variation under $B \to B+\d \lambda$ gives the following
  `t Hooft anomaly of the three-dimensional bosonic theory,
\begin{equation}
\label{3d anomaly with cup}
  \delta \mathcal{A} = i\pi \int_Y \delta (B\cup B) =
  i\pi \int_X B \cup \lambda + \lambda \cup B + \lambda \cup \d \lambda.
\end{equation}
Equation \eqref{Z_2^(d-2) anomaly} is the anomaly already considered in
\cite{3dbosonization2024}. Given its universality,
a simple way to check it is to look for a theory with
the desired properties. An example is the toric code, which is described
by the BF action $S= k/2\pi \int \z\d a$ involving two $\U(1)$ fields
$\z$ and $a$ and coupling constant $k=2$
\cite{gaiottokapustinspinTQFT1,gaiottokapustinspinTQFT2,mcgreevy2023generalized,luo2023lecture}. 
Indeed, this theory has a $\Z_2^{(1)}$ symmetry generated by a fermionic
quasiparticle (the product of the two Wilson loops of $\z$ and $a$)
with the above anomaly. This is described in detail in our previous work
\cite{3dbosonization2024}, and will be reviewed in Section
\ref{loop model sec}.

Let us mention that the anomaly \eqref{3d anomaly with cup} can be
rewritten in terms of more standard Abelian differential forms and
wedge product. In doing so, we employ the
notation of \cite{kapustinseiberg2014}: a $\Z_n$ gauge field $B$ can be
represented by a continuous $\U(1)$ gauge field $\hat{B}$ such that
$\d \hat{B}=0$ and its periods are $\Z_n$ valued,
\begin{equation}
  e^{i \oint \hat{B}} \in \Z_n \qquad \Rightarrow
  \qquad \oint \hat{B} = \frac{2\pi}{n}\Z.
\end{equation}
This amounts to the rescaling  
\begin{equation}
    B \to \hat{B} = \frac{2\pi}{n} B.
\end{equation}
Rewriting the three dimensional anomaly in terms of $\hat{B}$ yields
\begin{equation}
\label{3d bosonization anomaly}
\A = \frac{i}{\pi} \int_Y \hat{B} \wedge \hat{B}, \qquad \delta \mathcal{A}
= \frac{i}{\pi}\int_Y \delta (\hat{B} \wedge \hat{B}) =
\frac{i}{\pi}\int_X 2 \hat{\lambda} \wedge \hat{B} +
\hat{\lambda} \wedge \d \hat{\lambda}.
\end{equation}
This is the anomaly to be used in Section \ref{loop model sec} for
fermionization of the loop model in three-dimensions.  Hereafter we
stick to the continuous notation, dropping the hat and most of the
time also the wedge symbol.\footnote{Note that \eqref{3d bosonization
    anomaly} differs from \eqref{3d anomaly with cup} because $\cup$
  is not graded commutative. This fact has no consequences in the
  following analyses.}

%-3-----------------------------------------------

\section{Bosonization by flux attachment}
\label{fluxattachmentSEC}

The flux attachment is obtained by coupling
to a `statistical' gauge field with Chern--Simons action, 
which gives a magnetic flux to each particle proportional to its
charge (hence the name) \cite{cappellianomaliescondmat,marino}.
When this coupling is realized in relativistic field 
theory, it implies a web of infrared dualities
between 2+1d critical theories
\cite{tongwebofduality,wittenwebofduality}.
Broadly speaking, relativistic flux attachment is a map
between three-dimensional theories with $\U(1)$ (zero-form)
symmetry. Consider a theory $Z[A]$ with $\U(1)$ symmetry and
background gauge field $A$, then flux attachment consists in gauging
this symmetry by making $A$ dynamical and coupling to
the Chern--Simons (CS) theory:
\begin{equation}
\label{fluxattachment}
Z[A] \to \Tilde{Z}[A] = \int \mathcal{D}a Z[a]
\exp\left[i\int \frac{1}{4\pi}a\d a+\frac{1}{2\pi}a\d A \right].
\end{equation}
This procedure is consistent in three-dimensions since gauging an
$\U(1)$ symmetry produces a dual $\U(1)$ symmetry with conserved
current $*\d a/2\pi$ (the magnetic symmetry of the gauged theory). In
\eqref{fluxattachment} this current is then coupled
to $A$ again. As usual, background fields are convenient
tools to keep track of $\U(1)$ symmetries and their currents.

%-3.1--------------------------------------
\subsection{$\U(1)$ Chern--Simons theory is fermionic} 
\label{U(1) CS is spin}

The flux attachment \eqref{fluxattachment} transforms a
bosonic theory into a fermionic one. Besides the change in statistics
mentioned before, this is due to the fact that the Chern-Simons
theory with odd values of coupling constant $k$
(the so-called level)
depends on a spin structure, it is itself a fermionic theory. This is a
well--known fact
\cite{dijkgraafwitten,Witten2003SL2Z,seibergwitten2016gappedTI,
  wittenwebofduality}. Therefore,
if $Z$ in \eqref{fluxattachment} is a bosonic theory, after flux
attachment $\Tilde{Z}$ is fermionic.

Given its importance, let us review this result.  The fermionic nature
of $\U(1)_k$ CS theory\footnote{ A Chern--Simons theory with gauge
  group $G$ and level $k$ will be referred to as a $G_k$ CS theory.}
is rather clear from the physical point of view, because it appears in
low-energy effective theories of many fermionic systems in 2+1d and
1+1d. For example, it accounts for the gapped non-trivial phase of
electrons in the fractional quantum Hall effect. On its boundary, it
also predicts the presence of massless excitations, which involve a
chiral 1+1d fermion \cite{cappellianomaliescondmat}.

More technical arguments show the explicit dependence on the spin
structure of the apparently bosonic Chern-Simons theory. Let us write its
action,
\begin{equation}
\label{CS naive 3d}
    S = \frac{ik}{4\pi} \int_X a\wedge \d a,
\end{equation}
with $a$ a dynamical $\U(1)$ connection and $X$ an oriented
three-manifold, which also has a spin structure. When the gauge field
is topologically non-trivial, $f=\d a$ is globally well-defined, but
$a$ and the action are not.  To avoid this issue, a more proper
definition for the action is achieved by extending $X$ to a
four-dimensional manifold $Y$, such that $\p Y = X$, and define
\begin{equation}
\label{CS in 4d}
    S = \frac{ik}{4\pi}\int_Y f \wedge f, \qquad f = \d a.
\end{equation} 
The theory is still three dimensional if it does not depend
on the extension needed to define the action \eqref{CS in 4d}
\cite{wittenwebofduality}. By considering two possible extensions to
$Y$ and $Y'$, the difference between the two actions $I=S-S'$
is proportional to the instanton number for the closed 3+1d manifold
$Z= Y-Y'$, which takes the following values
\begin{equation}
  I \equiv S-S' = i\pi k \int_Z \frac{f}{2\pi} \wedge \frac{f}{2\pi} =
\begin{cases} 2\pi k\, \Z, & \text{if $Z$ is spin,}\\
           \pi k\, \Z, & \text{if $Z$ is not spin.}
\end{cases}
\end{equation}
We see that this expression depends on whether the 3+1d manifold $Z$
possesses a spin structure or not. If $k$ is an even
integer, the values of $I$ are in any case 
multiple of $2\pi$, i.e. are not observable.
Thus the extension of the Chern-Simons action is possible and tells us that it is well-defined independently of the spin structure. For odd $k$, $Y$
should be a spin manifold, thus the original 2+1d action is
actually sensible to the spin structure.  Finally, non-integer $k$ are not
allowed.\footnote{Actually, in this case
  the action \eqref{CS naive 3d} is also not gauge invariant
  on any $X$ \cite{witten2016threelectures}.}
Therefore, this argument shows that the $\U(1)_k$ CS theory is a
spin (fermionic) TQFT when $k$ is odd and bosonic when $k$ is even.

A more elaborate argument is needed to find the actual dependence of
the action \eqref{CS in 4d} and path integral on
the spin structure of $X$. Using the isomorphism
$\U(1) \cong \SO(2)$, we can consider the equivalent $\SO(2)_k$ CS. In
the path integral, one should sum over all $\SO(2)$ gauge bundles, which
are characterized by the values of $w_2(\SO) \in H^2(X, \Z_2)$, the second Stiefel-Whitney class for these bundles.
From the $\U(1)$ line bundle point
of view, $w_2(\SO)$ is the reduction mod 2 of the first Chern class
$\d a/2\pi$ \cite{CordovaHsinSeiberg2018SON(n)CS}: $w_2(\SO)$ is thus
dynamical and, being closed, it generates a $\Z_2^{(0)}$
symmetry by the topological defects
\begin{equation}
    (-1)^{\int_\Sigma w_2(\SO)}, \qquad \p \Sigma = 0.
\end{equation}
We can turn on a one-form background field $s$ for this symmetry, by the
coupling $S \to S+ ik\pi\int w_2(\SO) \cup s$. It can be shown that this
coupling changes the spin structure of $X$ by 
$\eta \ra \eta'= \eta+s$ \cite{jenquin2005spinCS,
  CordovaHsinSeiberg2018SON(n)CS}. Therefore, turning on $s$ is
equivalent to consider the CS theory with a different spin structure
$\eta'$,
\begin{equation}
\label{CS depends on spin}
S[\eta'] - S[\eta] = S[\eta+s]-S[\eta] = ik\pi \int_X w_2 \cup s =
ik \pi \{0,1\}.
\end{equation}
We thus found that $S \ra S + k\pi$ under changes of spin structure:
for $k$ odd the partition function gets a minus sign.
This finally determines how the CS theory depends on the spin
structure. This argument also shows that the symmetry
generated by $w_2(\SO)= \d a/2\pi$ (mod 2) is the fermion parity
symmetry $\Z_2^f$. This fact will be important in Section \ref{Bosonization
  of FA sec}.

We finally remark that for $k=1$, $\U(1)_1$ CS theory defines a
fermionic SPT phase: its Hilbert space is
one-dimensional (no topological order) and its partition
function is a complex number of modulus one
\cite{Witten2003SL2Z,wittenwebofduality}. This theory has only two Wilson line
observables, the identity and $W= e^{i\oint a}$. Being a spin TQFT, it
follows that $W$ represents a fermion.

%-3.2-------------------------------------------

\subsection{Boson-fermion duality by flux attachment}

Owing to the fermionic nature of $\U(1)_1$ CS theory, the flux
attachment \eqref{fluxattachment} is a kind of fermionization
procedure that turns a bosonic theory into a fermionic one. One could
wander whether it is possible to rewrite the resulting gauged theory in
terms of fermionic fields explicitly, making clear its nature. This is
not always possible (at least practically), but there are cases in
which the fermionic field description is known. 
The best example is given by the following IR duality
\cite{wittenwebofduality,tongwebofduality}:
\begin{equation}
\label{dirac = boson + FA}
\Bar{\psi} i \cancel{D}_A \psi + \left\{ -\frac{i}{8\pi} A\d A \right\}\qquad
\longleftrightarrow \qquad |D_a\phi|^2 + V(|\phi|)+ \frac{i}{4\pi} a\d a +
\frac{i}{2\pi} a \d A.
\end{equation} 
The left hand side is a free Dirac fermion coupled to the background
field $A$.  On the right hand
side, the scalar is coupled to a dynamical $\U(1)$ gauge field $a$,
which is then coupled to $A$, according to the flux attachment
\eqref{fluxattachment}. The potential term $V(|\phi|)$
tunes the scalar field to the Wilson-Fisher infrared fixed
point, where this duality holds. Notice that the
right hand side seems interacting, but, according to this duality, it
is in fact a free theory.

We remark that in the lhs of \eqref{dirac = boson + FA}
the fermionic theory is not
completely determined by the action, due to the parity anomaly, whose
value depends on the choice of regularization in the partition
function $Z$. The expression of the anomaly in curly brackets,
the $k=1/2$ CS action,
is valid for the definition $Z =|Z|e^{-i\frac{\pi}{2}\eta}$, with
$\eta$ the APS eta invariant of the Dirac operator. This term
is kept implicit in the definition of the Dirac operator in Ref.
\cite{witten2016fermion,wittenwebofduality},
while it is explicitly written in \cite{tongwebofduality}.
In the following, we shall adopt the first convention, omitting the
curly bracket in \eqref{dirac = boson + FA}.

The map \eqref{dirac = boson + FA} is actually a particular case
of the more general Chern-Simons plus matter dualities.
By assuming it, other dualities can be obtained, in particular
the boson-boson and fermion-fermion particle-vortex dualities
\cite{wittenwebofduality,tongwebofduality}. The relation \eqref{dirac =
  boson + FA} is itself a kind of particle-vortex duality: in
the bosonic theory, the fermion is expressed in terms of
the monopole $H(x)$, a solitonic excitation, in the form
$\psi\sim \phi^\dagger H$. Indeed, $\phi$ and $H$ have $\U(1)_a$ charges +1,
which cancels out, and $H$ has $\U(1)_A$ electric charge  $q=1$ and
unit  magnetic flux $\Phi = \int F = 2\pi$, leading to
$2\pi$ monodromy of the composite field with itself, and $\pi$
exchange phase, as required for fermions \cite{wittenwebofduality}.

The duality \eqref{dirac = boson + FA} is usually referred to as
three-dimensional bosonization \cite{tongwebofduality,
  3ddualityreviewSenthilSon}, as outlined in the Introduction.  This is
a consequence of the fundamental fields that appear in the two
theories. This terminology is also used for the non-Abelian
dualities involving fermions and scalars. However, according to the
analysis in the previous Sections, both theories are actually
fermionic, since they both depend on the spin structure.\footnote{Note
  that the relation \eqref{dirac = boson + FA} also holds in presence
  of a more general spin$_c$ structure, with $A$ a spin$_c$ connection
  \cite{wittenwebofduality} (see Appendix \ref{appBosonizationSpinc}
  for details). However, since a bosonic theory does not require
  neither a spin nor a spin$_c$ structure to be defined, the argument is
  the same. One is free to think to $A$ as a spin$_c$ connection to be
  more general.}  Therefore, the flux attachment \eqref{dirac = boson
  + FA} is more a non-trivial relation between two fermionic theories,
then a bosonization in the sense of the $\Z_2^f$ gauging
\eqref{invertible bosonization}. We are thus facing the two concepts
of bosonization in 2+1 dimensions which were alluded to in the
Introduction.  Beside matters of wording, the difference between these two
approaches has an important consequence: it is possible to gauge
fermion parity on both sides of \eqref{dirac = boson + FA} to get a
new kind of duality, as it will be shown in Section
\ref{Bosonization of FA sec}.

%-3.3-----------------------------
\subsection{Abelian dualities in 2+1 dimensions and SL$(2,\Z)$ action}
\label{Abelian dual sec}
The  boson and fermion particle-vortex dualities have the form
\cite{wittenwebofduality}, respectively,
\begin{equation}
\label{bb duality}
|D_A\phi|^2 + V(|\phi|) \qquad \longleftrightarrow \qquad
|D_a \Tilde{\phi}|^2 + V(\Tilde{|\phi|})+ \frac{i}{2\pi} a \d A \,,
\end{equation}
and 
\begin{equation}
\label{ff duality}
\Bar{\psi} i \cancel{D}_A \psi \qquad \longleftrightarrow
\qquad \Bar{\chi} i \cancel{D}_a \chi + \frac{i}{2\pi} a \d b -
\frac{2i}{4\pi} b \d b + \frac{i}{2\pi} b \d A - \frac{i}{4\pi} A \d A.
\end{equation}
These two maps can be derived from the
boson-fermion duality \eqref{dirac = boson + FA}
by imposing the correct action of time reversal on both its sides
\cite{3ddualityreviewSenthilSon}.  The
bosonic particle-vortex duality \eqref{bb duality} is the oldest and
most solid three dimensional duality. The scalars are always tuned to
the WF fixed point. The particle-vortex duality for fermions \eqref{ff
  duality} is more recent \cite{metlitski2016,Son-composite-fermion}. The
regularization for the fermion is implicitly assumed as said.  Note
that Eqs. \eqref{bb duality} and \eqref{ff duality} are
particle-vortex relations since the $\U(1)$ electric charge symmetry
on the lhs is mapped to the $\U(1)$ magnetic symmetry generated by
$*\d a$ on the rhs (thus $\phi$ ($\psi$) is dual to the monopole
operator in the theory $\Tilde{\phi}$ ($\chi$)).

The fermion-fermion duality \eqref{ff duality} is more involved than
its bosonic counterpart \eqref{bb duality}. It is possible to
simplify it by integrating out the dynamical $b$
field, leading to
\begin{equation}
\label{ff duality wrong}
\Bar{\psi} i \cancel{D}_A \psi + \frac{i}{8\pi}A\d A \quad
\longleftrightarrow \quad \Bar{\chi} i \cancel{D}_a \chi +
\frac{i}{8\pi} a \d a +  \frac{i}{4\pi} a \d A.
\end{equation}  
This is the original version of the fermionic duality as appeared in
\cite{metlitski2016}. Apart from the half-level Chern-Simons terms appearing on
both sides to cancel the respective anomalies, the relation
\eqref{ff duality wrong} is the same as the bosonic duality
\eqref{bb duality}, with one-half factor in the
term $a\d A$. As explained in \cite{wittenwebofduality} this gauge coupling is not
consistent with global flux quantization, but is valid on a local patch.
Therefore, \eqref{ff duality wrong} can be considered as an
approximate local version of \eqref{ff duality}.

All three dualities can be formulated in terms of two elementary maps
that correspond to the generators $S$ and $T$ of the $\SL(2,\Z)$ group
\cite{Witten2003SL2Z}. They are defined by:
\begin{equation}
\label{SL2Z generators}
\begin{split}
  &S: \qquad Z[A] \to \Tilde{Z}[A] =
  \int \mathcal{D}a Z[a] e^{\int \frac{i}{2\pi}a\d A};\\
    &T: \qquad Z[A] \to \Tilde{Z}[A] = Z[A]e^{\int \frac{i}{4\pi}A\d A}.
\end{split}
\end{equation}
They indeed satisfy $S^2 = -1$ (i.e. $J\to -J$ under $S^2$, which is
the original theory up to charge conjugation) and $(ST)^3=1$. The $T$
operation shifts the level by one in the Chern-Simons action, while the $S$ operation corresponds to gauging the $\U(1)$
symmetry.

The reformulation of 2+1d dualities in terms of 
$S$ and $T$ is as follows \cite{wittenwebofduality,
  3ddualityreviewSenthilSon}. The flux attachment \eqref{dirac =
  boson + FA} (boson-fermion duality) can be expressed as 
\begin{equation}
\label{dirac = FA with ST}
\vert\text{free Dirac fermion}\rangle =
ST\vert\text{scalar at Wilson-Fisher fixed point}\rangle.
\end{equation}
 The bosonic particle-vortex duality \eqref{bb duality} is
\begin{equation} 
\label{bb duality with S}
\vert\text{dual scalar  WF}\rangle =
S\vert\text{scalar WF}\rangle, 
\end{equation}
while the fermionic particle-vortex duality \eqref{ff duality} is
\begin{equation}
\label{ff duality with TS}
\vert\text{dual free Dirac}\rangle =
T^{-1} S^{-1} T^{-2} S^{-1} \vert\text{free Dirac}\rangle.
\end{equation}

We finally remark that the time reversal symmetry $\T$ reads, in
our notation,
\be
\T\vert\text{Dirac}\rangle \quad\to\quad
\vert \text{Dirac} + \frac{i}{4\pi} A\d A \rangle\,,
\ee
where the additional background term accounts for the anomaly
changing sign.

%-3.4---------------------------------------
\subsection{Comparing $\Z_2^f$ gauging with
  flux attachment in 1+1 dimensions }
\label{FA vs Z2 gauging in 2d}

The 2+1-dimensional bosonization/fermionization methods 
previously discussed, namely $\Z_2^f$ gauging
\eqref{fermionization} and flux attachment \eqref{fluxattachment},
have been shown to be rather different.
The fermionization map \eqref{fermionization}
can only be applied to bosonic theories with $\Z_2^{(1)}$ symmetry and
't Hooft anomaly \eqref{Z_2^(d-2) anomaly}, while the flux attachment
\eqref{fluxattachment} needs a $\U(1)$
symmetry.

A natural question is how these two methods can be compared with
well-established bosonization in $1+1$ dimension.  It can be shown that the $\Z_2^f$ gauging, being valid in any dimension,
reduces to standard 1+1d bosonization \cite{thorngren2020anomalies,tongArf,kapustinthorngren2017}. The flux attachment also has a
1+1d correspondent and to some extent it merges with the other method
\cite{tongArf,3ddualityreviewSenthilSon}. 

Within $\Z_2^f$ gauging, a simplification occurs because the 1+1d
anomaly \eqref{Z_2^(d-2) anomaly} for the $\Z_2$ (zero-form) symmetry
vanishes. It thus follows that any bosonic theory with $\Z_2$ symmetry
can be fermionized according to \eqref{fermionization}.  In Appendix
\ref{app gauge Z_2^f or spin str} we also remark that the term
$\int \eta\cup B$ can be represented in two dimensions by the Arf
invariant \cite{tongArf,backfiringbosonization2024}, which is the mod
2 index of the Dirac operator \cite{tongArf}. Thus, the $\Z^f_2$
gauging is equivalent to summing over $\eta$ in \eqref{fermionization}
with the introduction of the Arf invariant.
In Refs. \cite{tongArf,thorngren2020anomalies}, it is shown how this method reproduces known results
of 1+1d bosonization and extends them, e.g. by finding bosonic expressions
for partition functions in individual spin sectors.

Furthermore, since the
bosonic theory is not anomalous, the $\Z_2$ gauging can also be done
without including the Arf invariant (namely $\eta$), thus obtaining a
dual bosonic theory. As a result, it is possible to define two
operations $S$ and $T$ similar to \eqref{SL2Z generators}, that act on
1+1d theories with $\Z_2$ symmetry: $S$ is gauging the
$\Z_2$ symmetry and $T$ is the introduction of the Arf invariant. The
$\Z_2^f$ fermionization map \eqref{fermionization} is the combination $ST$.

After this identification, there is an analogy between the
$\Z_2^f$ gauging in 1+1d and the flux attachment in 2+1d. The $S$
operation consists in gauging the $\Z_2$ symmetry in 1+1d and the
$\U(1)$ symmetry in 2+1d. The $T$ operation introduces the Arf
invariant in 1+1d and the $\U(1)_1$ CS in 2+1d.
A duality web in two dimensions has been obtained in Ref.\cite{tongArf}, 
analogous to the three-dimensional case,
by acting with $S$ and $T$ on a seed
duality, which is the relation between a Majorana fermion
and the Ising model (which can be computed explicitly in two
dimensions). Notice, however, that $S$ and $T$ generate the different group
$\SL(2,\Z_2)$, since $T^2=1$.

In Ref. \cite{tongArf}
(cf. Appendix C) it was also argued that the Arf invariant arises from the
dimensional reduction of a $\U(1)_1$ CS to 1+1
dimensions. According to this result, three-dimensional flux
attachment becomes equivalent to $\Z_2^f$ gauging
in 1+1 dimensions.

%-4---------------------------------------------------
\section{Bosonizations and dualities in the loop model}
\label{loop model sec}

In this Section we study both the $\Z_2^{(1)}$ gauging
\eqref{fermionization} and the flux attachment \eqref{fluxattachment}
by explicitly evaluating them on the loop model. This is a
semiclassical, yet non-trivial conformal theory
\cite{loopmodel,3dbosonization2024}, which realizes the universal
quadratic response of three-dimensional CFTs with $\U(1)$ symmetry
\cite{Witten2003SL2Z} and displays a lot of interesting features. In
Refs.\cite{loopmodel,3dbosonization2024}, it was introduced as a theory for
massless degrees of freedom at the surface of 3+1d topological
insulators. Here we mainly focus on the theory itself, as an example
of bosonic theory to be made fermionic by using the two maps discussed
in the previous Sections.

\subsection{Introducing the loop model}

The loop model is defined by the action \cite{loopmodel,3dbosonization2024}:
\be
\label{loop-model}
  S[\zeta, a, A] = \frac{i}{2\pi}\int_X   k \zeta \d a +
    \zeta \d A  +   \frac{g_0}{4\pi} \int_X a_\mu
  \frac{-\delta_{\mu \nu} \partial^2 + \partial_\mu \partial_\nu}
  {\partial} a_\nu \,,
\ee
where $a$ and $\z$ are dynamical $\U(1)$ fields, $k\in
\Z$, and $A$ is the background electromagnetic field. Under time reversal, $a$ and $\z$ transform, respectively, as vector and pseudovector, such that the action is invariant. Furthermore,
$\p^2 = \p_\mu \p_\mu$ and $\partial^{-1}$ is the Green function of
$\partial \equiv \sqrt{-\partial^2}$ in three Euclidean
dimensions:
\begin{equation}
    \frac{1}{\p (x,y)}=\frac{1}{2 \pi^2} \frac{1}{(x-y)^2}.
  \end{equation}
The action involves the following kernel
\begin{eqnarray}
&&\frac{1}{4\pi}\int{d^{3}x\, d^{3}y\,
    a_{\mu}(x) D_{\mu \nu}(g,f) a_{\nu}}(y)\, ,
  \nonumber\\
&&  D_{\mu \nu}(g,f) =  g \dfrac{1}{\p} ( -\delta_{\mu \nu} \p^{2} +
  \p_{\mu} \p_{\nu}) +if \varepsilon_{\mu \rho \nu} \p_{\rho}\,,
  \label{S_loop_ker}
\end{eqnarray}
which satisfies an interesting inversion relation 
\begin{align}
& \int j_{\mu} D^{-1}(g,f)_{\mu \nu}j_{\nu} =
                \int \zeta_{\mu} D_{\mu \nu}(\wh{g}, \wh{f})  \zeta_{\nu}\, ,
\qquad j_{\mu}= \varepsilon_{\mu\nu\rho} \p_{\nu}\zeta_{\rho} \, ,
  \\
  & \t=f+ig, \qquad \wh \t=\frac{-1}{\t}
    \qquad\wh{g}= \dfrac{g}{g^{2}+f^{2}}, \qquad
  \wh{f}= \dfrac{-f}{g^{2}+f^{2}}\,.
\label{g_f_tilde}
\end{align}
This generalizes the expression in
\eqref{loop-model} by including a Chern-Simons term with coupling $f$.
The two coupling constants can be conveniently combined in the complex
number $\t=f+i g$. These formulas can be used to integrate $a$ in \eqref{loop-model}
and obtain
\be
\label{loop-model2}
S[\zeta, A] = \frac{i}{2\pi}\int_X   \zeta \d A  +
\frac{k^2}{4\pi g_0} \int_X \z_\mu
  \frac{-\delta_{\mu \nu} \partial^2 + \partial_\mu \partial_\nu}
  {\partial} \z_\nu \,.
\ee

In the physical setting of 3+1d topological insulators (fractional for
$k\neq 1$), the bulk theory is described by a BF topological theory \cite{BFTI},
which implies the 2+1d BF term $(k/2\pi)\int\z \d a$ in the loop
model action \eqref{loop-model}. The non-local term provides the
dynamics for the surface excitations.
When $k=1$, the electromagnetic response of \eqref{loop-model}
is indeed the one of a three-dimensional Dirac fermion at one
loop, which also dominates in the limit of large number of fermions
$N_f$.  Notice
that in this physical setting, the action \eqref{loop-model} should also
include the term $(1/8\pi) a\d A$
to reproduce the parity anomaly in the one-loop
fermionic response. This term is assumed to be canceled by the theta term
in the bulk, as usual in topological insulators \cite{3dbosonization2024}.

The theory \eqref{loop-model} has been studied in detail in
\cite{loopmodel,3dbosonization2024}, for general $k\neq 1$.
Despite its non-local nature, it
is a conformal theory at the quantum level, with a critical line
parameterized by the coupling $g_0$ \cite{motrunich2012,loopmodel}. It
corresponds to the large $N_f$ limit of mixed-dimensional QED, where
three-dimensional fermions interact with four-dimensional photons
\cite{sonmixedqed,sonmixedqedscalar}. In the following, we recall two main
properties of the theory \eqref{loop-model}, its partition function
on the conformal cylinder $\R\times S^1$, describing the conformal
spectrum, and its exact selfduality.

%-4.1.1------------------------------
\subsubsection{Partition function} 
The partition function of \eqref{loop-model} has been
computed on the
three-dimensional cylinder $S^2 \times \R$, with radius $R$  \cite{loopmodel}. It was obtained by rewriting the loop
model as a local theory in four dimensions (with the extra dimension
just a fictitious one), resolving the issues related to its nonlocal
nature. The result is:\footnote{The partition function is computed on
  the periodic time interval $L=\b=1/k_B T$, thus the geometry is
  actually $S^2 \times S^1$.}
\begin{equation}
\begin{split}
  \label{loop-model-Z}
  Z[S^2 \times S^1] &= Z_{osc}\, Z_{sol}= Z_{osc}\sum_{N_0,M_0 \in \Z}
  \exp \left( -\frac{\beta}{R} \Delta_{N_0,M_0} \right),
\\
   \Delta_{N_0,M_0} &= \frac{1}{2\pi} \left( \frac{N_0^2}{g_0} +
    g_0 \frac{M_0^2}{k^2} \right).  
\end{split}
\end{equation}
In this equation, $Z_{sol}$ is obtained by evaluating the classical
action on soliton configurations, leading to the spectrum of scaling
dimensions $\D_{N_0,M_0}$; $Z_{osc}$ describes fluctuations around
classical solutions (its explicit form is given in
\cite{loopmodel}). The integers $N_0$, $M_0$ parameterize the fluxes of
the gauge fields $\zeta$, $a$,
\begin{equation}
\label{boundary-charges}
Q= \frac{1}{2\pi}\int_{S^2} \d \zeta  = \frac{N_0}{k},
\qquad\qquad Q_T = \frac{1}{2 \pi} \int_{S^2} \d a  = \frac{ M_0}{k}.
\end{equation}
The corresponding solitons of the theory (\ref{loop-model}) are static
solutions of the equations of motion with boundary conditions suitable
for these charges. Note that the $\z$ flux characterizes a soliton
with electric charge $Q$, owing to the coupling to the electromagnetic
background in \eqref{loop-model}.
The $a$ fluxes correspond to magnetic charge $Q_T$.
Therefore, the loop model possesses conformal fields with dyonic
charges\footnote{
Note that the fractional quantization of the charges \eqref{boundary-charges}
is actually resolved by considering a bulk which carries the
opposite charged excitations, thus recovering integer-valued fluxes
for $a$ and $\z$ in the whole system.}
$(N_0,M_0)$, $N_0,M_0\in\Z$. Owing to the BF topological term in
\eqref{loop-model}, the monodromy phase between these excitations is
\be
\label{monod}
\frac{\th}{2\pi}=\frac{N_0 M_0}{k}\,.
\ee

%-4.1.2-------------------------
\subsubsection{Bosonic particle-vortex duality}

The loop model \eqref{loop-model} is particularly interesting because it
enjoys explicit selfduality.
The property can be seen in the solitonic spectrum
in \eqref{loop-model-Z} and the monodromies \eqref{monod}, which are invariant
under the exchanges
\begin{equation}
\label{loop-model-duality}
g_0\ \lra\ \wt g_o=\frac{k^2}{g_0}, \qquad  N_0 \ \leftrightarrow -\ M_0.
\end{equation}
The selfduality can be verified 
at the level of the action \eqref{loop-model}, by applying
the particle-vortex map
\eqref{bb duality} extended to $k\neq 1$, as follows:
\begin{equation}
\label{bosonic PV with k}
\wt Z[A]=\int {\cal D}c\, Z[k c] \exp\left(\frac{i}{2\pi}\int c\,  \d A\right).
\end{equation}
This amounts to replacing $A$ in the action \eqref{loop-model}
with the auxiliary dynamical field $kc$,
which is then coupled to $A$.  The integration over $c$ can
be done explicitly and the resulting expression
can be compared with \eqref{loop-model2}.  This reproduces
the map  between 
coupling constants \eqref{loop-model-duality} and between the fields
\begin{equation}
a \ \leftrightarrow \ -\zeta,
\end{equation}
implying those for the spectrum \eqref{loop-model-Z}
and monodromies \eqref{monod}.

%-4.1.3----------------------
\subsubsection{Loop model dynamics: universal quadratic response}

Consider a 2+1d conformal theory with a $\U(1)$ symmetry and
associated conserved current $J$. The two-point function of the
current is completely fixed by the two symmetries to be
\begin{equation}
  \braket{J_\mu (p)J_\nu(-p)} =
  \frac{g}{2\pi}\frac{\delta_{\mu\nu}p^2-p_\mu p_\nu}{p}+
  \frac{f}{2\pi} \epsilon_{\mu\nu\rho}p_\rho \,,
\end{equation}
where $g$, $f$ are two dimensionless constants and $p = \sqrt{p^2}$. We can
couple the theory to a background gauge field $A$ and consider the
generating functional of the current correlators $Z[A] =
e^{-\G[A]}$. By the above correlator, it follows that the effective
action in terms of $A$ at quadratic order is completely fixed by the
symmetries,
\begin{equation}
\label{loop-model response of 3dCFTs}
\G[A;g,t]= \int \frac{g}{4\pi} A_\mu \frac{-\delta_{\mu \nu} \partial^2 +
  \partial_\mu \partial_\nu}{\partial} A_\nu + \frac{if}{4\pi} A\d A + O(A^3).
\end{equation}
This is indeed the non-local electromagnetic response of the loop model
\eqref{loop-model} after integrating $\z$.  

There is one caveat for this result. In general, conformal field theories are
strongly coupled and the effective action
$\Gamma[A]$ cannot be truncated at quadratic order, as in
\eqref{loop-model response of 3dCFTs}. However, there are physically
significative cases where the effective action
is actually weakly coupled. 
An interesting example is given by the
large $N_f$ limit of mixed-dimension QED$_{4,3}$, involving 2+1d
fermions and 3+1d photons, which is exactly particle-vortex selfdual 
and possesses a critical line 
\cite{sonmixedqed,sonmixedqedscalar,dudal2019exact} like the loop model
\cite{loopmodel,motrunich2012}. In this case the
higher order terms in \eqref{loop-model response of 3dCFTs} are of
order $1/\sqrt{N_f}$, thus the theory described by $\G[A]$ makes sense.
More generally, this truncation corresponds to the semiclassical
limit $\hbar\to 0$, in which three- and higher-point functions of currents
are neglected. These could be added by a perturbative expansion in $1/\sqrt{N_f}$.

%-4.2---------------------------------------
\subsection{Fermionization of the loop model by $\Z_2^{(1)}$ gauging}
\label{loop-model Z2 gaug sec}
In this Section we review the fermionization map 
\eqref{fermionization} applied to the theory \eqref{loop-model}, which
was carried out in \cite{3dbosonization2024}.
We first recall the main steps of this map 
discussed in Section \ref{bosonizationSEC}. The bosonic theory dual to
a fermionic one should have a $\Z_2^{(1)}$ symmetry with 't Hooft
anomaly \eqref{3d bosonization anomaly}, expressed in terms of the
2-form $\Z_2$ background field $B$.
On one hand, the bosonic theory is obtained by gauging
$\Z_2^f$ in the fermionic theory and it is therefore a theory of fermion
bilinears. On the other hand, the fermionic states can be probed by using
the background $B$: in the $B$-twisted Hilbert
space, states are fermionic, since after gauging $\Z_2^{(1)}$ the generator of
fermion parity is indeed $e^{i\int B} = (-1)^F$. This allows to obtain
an invertible map between the spin sectors of the fermionic theory and
the $B$-twisted sectors of the bosonic theory, which is the explicit
version of \eqref{fermionization}. Finally, since $\T_f^2 = (-1)^F$ in
the fermionic theory, in the bosonic theory $\T_b^2 = e^{i\int B}$. In
the following, we will see all these aspects.

We start by discussing the topological features of the loop model,
which are determined by the BF term in the action \eqref{loop-model}
having coupling constant $k\in \Z$.
These features are not affected by the dynamic term,
with coupling $g_0$, because the equations of motion of the full action
still imply $\d a = \d \zeta = 0$ in vacuum.
In particular, the monodromy between
closed Wilson loops of the gauge fields is given by their
linking number:
\begin{equation}
  \braket{e^{in\oint_\gamma \zeta} e^{im\oint_{\gamma'} a}} =
  e^{\frac{2\pi i}{k}nm LN(\g,\g')}.
\end{equation}
Thus, the loop model possesses a
$\Z_k^{(1)} \times \Z_k^{(1)}$ symmetry.

Although the loop model arises in the context of fermionic topological
insulators, the BF theory is bosonic, i.e. it does not depend on the
spin structure of the manifold, contrary to the Chern-Simons
theory. This can be understood by rewriting the BF term as the
difference of two independent Chern-Simons theories with coupling
$2k$, which are bosonic.  Another aspect is that the spectrum
\eqref{loop-model-Z} for $k=1$ (no topological order) does not contain
a solitonic excitation with fermionic self-statistics. Therefore, the
bosonic BF theory does require fermionization for becoming a spin
TQFT. The interesting aspect is that this transformation, earlier
described at topological level, can now be made in presence of
dynamics.

In the previous work, the $\Z_2^{(1)}$ gauging
was applied to the loop model for two
values of the couplings, $k=1$ and $k=2$.  Each choice has advantages
and disadvantages.  The $k=1$ case is more natural because it follows
from effective descriptions of the fermionic topological insulators,
e.g. reproducing its parity anomaly. In this case, the $k=1$ bosonic
spectrum \eqref{loop-model-Z} should be first (selfconsistently)
deformed by allowing a state with charges $(N_0,M_0)=(1,1/2)$, the
would-be fermion, also realizing the needed $\Z_2^{(1)}$ (anomalous)
symmetry \cite{3dbosonization2024}.  The case $k=2$ may seem puzzling,
given that the bosonic spectrum has semion excitations with $Q=1/2$
and $Q_T=1/2$ and topological order.  Nonetheless, it naturally has
the $\Z_2^{(1)}$ symmetry and it is found that the unwanted semions
and topological order are eliminated from the final fermionic theory.
In the following we review the fermionization in the simpler $k=2$
case.

When $k=2$, the topological
BF theory is the low energy limit of the toric code \cite{mcgreevy2023generalized,luo2023lecture}. It has a
$\Z_2^{(1)} \times \Z_2^{(1)}$ symmetry whose diagonal subgroup,
generated by the fermionic quasiparticle
\begin{equation}
\label{fermion toric code}
   \psi = e^{i\oint a +\zeta},
\end{equation}
has indeed the 't Hooft anomaly \eqref{3d bosonization anomaly}. We
can therefore couple \eqref{loop-model} to the two-form background
gauge field $B$ for the diagonal symmetry $\Z_2^{(1)}$ \eqref{fermion
  toric code} as follows \cite{3dbosonization2024} (equation (4.69) there)
\begin{equation}
\label{k2-boundary-action}
  S =  \frac{i}{2\pi}\int_X 2 \zeta \d a +
  2 \left(a +\zeta \right) B +  2(\d \zeta
  +B) A  + \frac{g_0}{8\pi}\int
  \left(f +B \right)_{\mu\nu} \frac{1}{\p}  \left(f +B\right)^{\mu\nu}.
\end{equation}
Under the diagonal background gauge transformation 
\begin{equation}
    \zeta \to \zeta - \l, \qquad a \to a- \l, \qquad B \to B + \d \l,
\end{equation}
one finds that the action \eqref{k2-boundary-action} is not invariant by the anomaly
term \eqref{3d bosonization anomaly}, as anticipated. 

The action \eqref{k2-boundary-action} can be written in terms of the gauge invariant combinations $\d \Tilde{a} = \d a +B$, $\d \Tilde{\z} =\d \z + B$ and subtracted of the anomaly term, thus obtaining the original $B=0$ expression \eqref{loop-model} (see equation (4.37) in \cite{loopmodel}). It follows that the only role of the flat $\Z_2$-valued $B$ field is to
imposing constraints
on the global fluxes \eqref{boundary-charges}, without altering the
dynamics. This can be seen directly from the equations of motion for
\eqref{k2-boundary-action} (with $A=0$), $\d a + B=0$, $\d \z+B=0$,
which actually mean the following conditions on the
fluxes \eqref{boundary-charges}
\be
\label{k2-sectors}
\begin{split}
& - \frac{1}{2\pi} \int_{S^2} B =\frac{1}{2\pi}\int_{S^2} \d\z=Q=
  \frac{N_0}{2}=0,\frac{1}{2}, \quad {\rm mod}\ 1,
  \\
& - \frac{1}{2\pi} \int_{S^2} B =\frac{1}{2\pi}\int_{S^2} \d a =Q_T=
\frac{ M_0}{2}=0,\frac{1}{2},\quad {\rm mod}\ 1.
\end{split}
\ee
 The solutions, in
the geometry $X = S^2 \times \R$, identify two sectors: an even sector,
for $\int B = 0$ mod $2\pi$, with $N_0$ and $M_0$ even, and an odd
(twisted) sector, for $\int B = \pi$ mod $2\pi$, where $N_0$ and
$M_0$ are odd. In the gauged theory, $\exp(i \int_{S^2} B) = (-1)^F$,
thus these are the bosonic (even number of fermions)
and fermionic (odd number) sectors, respectively.

As described in Section \ref{bosonizationSEC}, the fermionic partition
function is obtained by gauging the $\Z_2^{(1)}$ symmetry, as in
\eqref{fermionization}. Concretely, the bosonic $(k=2)$ partition
function $Z_b[B]$ is summed over the $B$-field values
\begin{equation}
\label{Z_f-S}
Z_f[\eta+s] = \sum_{B \in H^2(X; \Z_2)} Z_b[B] e^{\frac{i}{\pi}\int_Y B\wedge B}
e^{\frac{i}{\pi} \int_X B \wedge s}.
\end{equation}
The $s$ background field is an element of $H^1(X,\Z_2)$, which couples
to $B$ in order to shift from one spin structure to another in the
fermionic theory. Therefore,
\eqref{Z_f-S} is precisely of the form \eqref{fermionization} with spin
structure $\eta +s$.

On $S^2\times S^1$ there are two spin structure $\eta_\pm$, for
antiperiodic ($NS$ sector) and periodic ($\wt{NS}$ sector) boundary
conditions for the fermion on the time circle. The field $s$ in \eqref{Z_f-S}
allows to shift $\eta_+$ to $\eta_+ + s = \eta_-$ (and
viceversa). Fixing $\eta_+$, say, then $\eta_-$ is obtained by
inserting a flux of $s$ along the time circle, $\int_{S^1} s =
\pi$. The two possible fermionic partition functions are therefore
\begin{equation}
\label{Z_f}
Z_f^{NS}= Z_f[\eta_+; {\textstyle\int_{S^1}} s =0], \qquad\qquad
Z_f^{\widetilde{NS}}=Z_f[\eta_+, {\textstyle\int_{S^1}} s = \pi]\,.
\end{equation}
To write these explicitly, we consider the bosonic partition functions
\eqref{loop-model-Z}, in which the $B$ field selects the
two sectors \eqref{k2-sectors} (the anomaly has been removed from the
action \eqref{k2-boundary-action}). Their expressions are
\begin{equation}
\label{Z_B}
\begin{split}
  &Z_b[{\textstyle\int} B=0] = Z_{osc}\sum_{(N_0,M_0)  \in 2\Z}
  e^{-\frac{\beta}{2 \pi R} \left(\frac{N_0^2}{g_0} + g_0 \frac{M_0^2}{4} \right)},
  \\
  &Z_b[{\textstyle\int} B=\pi] =Z_{osc} \sum_{(N_0,M_0) \in 2\Z + 1}
  e^{-\frac{\beta}{2 \pi R} \left(\frac{N_0^2}{g_0} + g_0 \frac{M_0^2}{4} \right)}.
\end{split}
\end{equation}
Then the fermionic partition functions \eqref{Z_f} are 
\begin{equation}
\label{Z_f-NS}
Z_f^{NS} =  Z_b[0] + Z_b[\pi], \qquad
Z_f^{\widetilde{NS}} =  Z_b[0] - Z_b[\pi],
\end{equation}
and their bosonic inverses \eqref{invertible bosonization} are 
\begin{equation}
\label{Z_b-sum}
Z_b[0] = \frac{1}{2}(   Z_f^{NS} + Z_f^{\widetilde{NS}}),
\qquad\qquad Z_b[\pi] =  \frac{1}{2}( Z_f^{NS} - Z_f^{\widetilde{NS}}).
\end{equation}
These expressions show how bosonization by $\Z_2^f$ gauging, and its inverse,
fermionization by $\Z_1^{(1)}$ gauging, are realized
in the loop model, in agreement with
\eqref{invertible bosonization} and \eqref{fermionization}.
 
Notice that the bosonic partition function $Z_b[0]$
can be written in Hamiltonian formulation as follows
\begin{equation}
  \label{Z_B0}
    Z_b[0] =   \frac{1}{2} \text{Tr}\left[ (1 + (-1)^F)e^{-\beta H} \right],
\end{equation}
where the trace is taken over the fermionic Hilbert space on the
spatial $S^2$. The factor $(-1)^F$ occurs for periodic boundary
conditions in the time direction $S^1$. The expression
(\ref{Z_B0}) only involves
states with an even number of fermions, which are then bosonic,
as expected.

The introduction of the $B$ field allows also to correctly reproduce
the characteristic fermionic feature $\mathcal{T}^2 = (-1)^F$. As
said, we have that $(-1)^F = e^{i\int_{S^2}B}$ and it is natural to
write in the bosonic theory
\begin{equation}
\label{T-square}
    {\cal T}^2 = e^{i\int_{S^2}B}.
\end{equation}
The action of time reversal on states is defined to be consistent with
\eqref{T-square}. The physical states in the pure gauge theory
\eqref{k2-boundary-action} are created by the Wilson lines of the
gauge fields $a$ and $\zeta$,
\begin{equation}
  \label{gen-stat}
\Phi = e^{-i n \int a-i m\int \z}.
\end{equation}
We add a non-trivial $B$-dependent phase factor to the transformation,
\be
\label{T-states}
  {\cal T} \Phi  {\cal T}^{-1} \equiv e^{i\int_{S^2}\frac{B}{2}} \,
e^{-i n \int a +i m \int \z} \,,
\ee
such that
\begin{equation}
\label{T2-odd}
{\cal T}^2\Phi  {\cal T}^{-2}={\cal T}\left(e^{i\int_{S^2}\frac{B}{2}}
  e^{-in\int a +i m \int \z}\right) {\cal T}^{-1}=
  e^{i\int_{S^2}B} \Phi,
\end{equation}
which becomes $(-1)^F \Phi$ after gauging. Notice that the phase is
trivial, $\int B = 0$, for all bosonic states in the even sector of
the theory, thus ${\cal T}^2=1$ on them, while in the odd fermionic
sector, $\int B = \pi$ and ${\cal T}^2=-1$ correctly.

The $B$ field modifies the standard $\T^2= 1$
condition valid for bosonic theories.
Equation \eqref{T-square} is therefore a signal of a mixed anomaly
between the $\Z_2^{(1)}$ symmetry and $\T$ in the bosonic
theory, which can be probed by considering the theory on non-orientable
manifolds. This result is explicitly derived at the end of Appendix
\ref{appBosonizationSpinc}. If we want to define the fermionic
theory on a non-orientable manifold, there is an extra anomaly piece to be
removed besides \eqref{3d bosonization anomaly} (given in
\eqref{bosonization anomaly with TR}).

%-4.2.1-----------------------------
\subsubsection{Duality of the fermionic spectrum}
A final remark concerns the fate of the selfduality obeyed by the bosonic 
loop model \eqref{bosonic PV with k} after fermionization.
The spectrum of the fermionic partition function $Z_f[\eta_\pm]$
\eqref{Z_f} is still invariant under the $k=2$ bosonic duality
\eqref{loop-model-duality}, 
\begin{equation}
\label{fermionic loop model duality}
    g_0\ \lra\ \frac{4}{g_0}, \qquad N_0 \ \lra\  M_0,
\end{equation}
because the parity rule $N_0=M_0$ mod $2$ in \eqref{Z_B} also respects it.
However, the duality map \eqref{bosonic PV with k} should be modified because 
the electric charge for excitations is
twice the value in \eqref{boundary-charges}, as needed for the
unit charge of the fermion $(N_0,M_0)=(1,1)$. The duality now consists in
replacing the $A$ background with the dynamical field $c$ and adding a
BF coupling $2c \d A /2\pi$ to \eqref{k2-boundary-action}.
In conclusion, this duality is specific of the bosonic loop model
and is only mildly affected by $\Z_2^f$ gauging. This should be
contrasted with the strong interplay between flux attachment
\eqref{dirac = boson + FA} and particle-vortex
maps \eqref{bb duality}, \eqref{ff duality} discussed 
in Section \ref{Abelian dual sec}.

%-4.3---------------------------------------------------
\subsection{Fermionization of the loop model by flux attachment}

We start by recalling the loop model action \eqref{loop-model}
\be
\label{b-loop}
  S_b=\frac{g_0}{4\pi}\int a D a +\frac{i}{2\pi}\int k \z \d a + \z\d A \,,
\ee
where the short-hand notation,
$aDa= a_\mu ((\p_\mu \p_\nu -\delta_{\mu\nu} \p^2)/\p)a_\nu$, is
used for the non-local kernel.
The flux attachment \eqref{dirac = boson + FA} is introduced in the following
generalized form for $k\neq 1$
\be
\label{loop-FA}
S_b\quad\to\quad
S_f=\frac{g_0}{4\pi}\int a D a + \frac{i}{2\pi}\int k \z \d a + 
\left(k \z\d c+ \frac{k}{2} c\d c +c \d A \right)
      +\frac{1}{4k} A\d A \,.
\ee
In this expression, the coupling to the field $c_\mu$ is
defined by  $A_\mu \to k c_\mu $, as done in the bosonic
particle-vortex duality \eqref{bosonic PV with k}.  The other terms are the
Chern-Simons action (the level $k$ is assumed odd) and the parity
anomaly $1/4k A\d A$ (in curly brackets in \eqref{dirac = boson +
  FA}), which must be explicitly included in effective field theory
descriptions. Its coefficient is found by assuming anomaly inflow from the bulk
of 3+1-dimensional (fractional) topological insulators (see Eq. (2.3) in
\cite{3dbosonization2024}), which can be a physical realization of this model.

Integration over $c_\mu$ gives the expression
\be
\label{f-loop}
S_f[a,\z,A]=\frac{g_0}{4\pi} \int a D a
+\frac{i}{2\pi}\int k \z \d a -\frac{k}{2} \z\d\z-\z\d A -
\frac{1}{4k} A\d A \,.
\ee

Setting temporarily the background $A=0$, we can discuss some basic
features of this theory. We first recall the equations of motion of
the bosonic model \eqref{loop-model}, for comparison,
\be
\label{b-loop-eq}
\d a=0, \qquad *\d\z -i\frac{g_0}{k} Da=0 \,.
\ee
These imply $\d a=\d\z=0$ and thus the absence of local
fluctuations in vacuum.
The only possible excitations are solitonic, corresponding to singular
solutions $\d a\neq 0$ for $x=x_0$, basically determined by the topological
part of the theory (corresponding to the $g_0=0$ limit)
\cite{3dbosonization2024}.

The equations of motion in the flux-attached theory are rather different,
\be
\label{f-loop-eq}
\d a=\d\z, \qquad *\d a -i\frac{g_0}{k} Da=0, 
\ee
because they allow for vacuum excitations, although not of ordinary photons.
Therefore, the Chern-Simons action introduces dynamics in the theory
which makes it rather different from its topological limit $g_0=0$.
Nonetheless, we shall see that it is possible to obtain the solitonic spectrum
and correspondingly the
solitonic part of the partition function on the geometry $S^2\times S^1$
(cf. Section 4.1.1), within the semiclassical limit.

It is interesting to integrate out the field $\z$ from the action
\eqref{f-loop} to obtain the reduced expression
\be
\label{f-loop-a}
S_f[a,A]=\frac{g_0}{4\pi}\int a D a +\frac{i}{2\pi}\int k a\d a
-a\d A +\frac{1}{4k}A\d A  \,.
\ee
In alternative, we can eliminate $a$ in \eqref{f-loop}
by using the loop-model identity
\eqref{g_f_tilde}, thus finding another reduced action
\be
\label{f-loop-z}
S_f[\z,A]=\frac{k^2}{4\pi g_0}\int \z D\z  +\frac{i}{2\pi}\int - k \z \d \z
-\z\d A -\frac{1}{4k} A\d A\,.
\ee
These two expressions show that  the gauge fields $a,\z$
play an equivalent role in this theory, with detailed aspects to be
clarified in the following discussion.

%-4.3.1------------------------------------
\subsubsection{Partition function}

In the bosonic theory, the spectrum was obtained by evaluating the
action, suitably extended in one extra dimension, for smooth classical
solutions corresponding to the fluxes of the fields $a,\z$
\eqref{boundary-charges} in the $S^2$ spatial geometry (cf. Section 4
of Ref. \cite{loopmodel}).  This analysis extends to the present case,
where it is apparent that the equations of motion \eqref{f-loop-eq}
identify electric and magnetic charges.  The closer inspection of the
solutions of equation of motion for the action $S[a, 0]$ ($A=0$)
\eqref{f-loop-a}, shows that $\int_{S^2} \d a\neq 0$ corresponds to a
smooth magnetic field but also provides the source for the electric
field, with no conditions relating the two parts.

The result is the following spectrum for the fermionic theory
\be
\label{f-spec}
N_0=M_0, \qquad \qquad \D^f_{N_0=M_0}=\frac{1}{2\pi}N^2_0
\left(\frac{1}{g_0}+\frac{g_0}{k^2} \right) ,
\ee
which confirms the naive expectation that the Chern-Simons
theory attaches magnetic flux $q/k$ to excitations with charge $q$.
Furthermore, it is natural that the solitons with $N_0=M_0$ have self-monodromy
\be
\label{f-monod}
\frac{\th}{2\pi}=\frac{N^2_0}{k} \,, \ee whose half value gives the
fermionic statistics for $N_0^2/k=$ odd (in particular $N_0=k=1$).
This result will be confirmed later by evaluating correlation
functions.

The partition function of the flux-attached loop model \eqref{loop-FA}
is therefore
\be
\label{Z-f-loop}
Z_f^{NS}=Z_{osc} \sum_{N_0\in\Z} \exp\left( -\frac{\b}{2\pi
    R}N_0^2\left( \frac{1}{g_0}+\frac{g_0}{k^2} \right) \right).
\ee
Note that the spectrum is invariant under
$g_0 \leftrightarrow k^2/g_0$, which corresponds to the exchange
$a\leftrightarrow \z$ in the actions \eqref{f-loop-a} and
\eqref{f-loop-z}, combined with time-reversal symmetry.  This symmetry
was also found in the bosonic model by particle-vortex duality, while
here it is realized with same Lagrangian.

Summarizing the discussion of fermionization of the
loop model, the following spectra were obtained by $\Z_2^f$ gauging
and flux attachment:
\begin{eqnarray} 
\label{spectra}
&&  Q=\int\frac{\d\z}{2\pi} =N_0, \qquad Q_T=\int\frac{\d a}{2\pi}=M_0,
     \qquad
     \D_{N_0,M_0}=\frac{1}{2\pi}
\left(\frac{N_0^2}{g_0}+g_0 M_0^2 \right),
     \nonumber\\
  && \text{bosonic:}\qquad\qquad\quad\ (N_0,M_0)=(n,m), \quad n,m \in\Z,
     \nonumber\\
  && \Z_2^f\ \text{gauging:}\qquad\qquad (N_0,M_0)=(n,m/2), \quad n,m \in\Z,
     \quad n=m\ {\rm mod}\ 2,
     \nonumber\\
&&  \text{flux attachment:}\qquad (N_0,M_0)=(n,n),\quad n \in\Z,
\end{eqnarray}
having set $k$ to proper values (respectively, $k=1,2,1$).
Note that the basic fermion excitations have charges $(1,1/2)$ and
$(1,1)$ in the two approaches.

Therefore, we have explicitly checked that the two 
bosonizations in 2+1 dimensions yield different fermionic theories
when applied to the loop model.

%-4.3.2--------------------------------------
\subsubsection{Fermionic fields and correlators}

We already pointed out that the topological aspects of the bosonic loop model
are not hampered by dynamics, which keeps the connections flat,
$\d a=\d \z=0$ in \eqref{b-loop-eq}.
It follows that solitonic excitations can be represented by Wilson loops
of the gauge fields, as in the topological limit
(see \eqref{fermion toric code}):
in particular, the endpoints of open loop generate 
localized values of $Q,Q_T$ \eqref{b-loop-eq}.
These properties extend to the fermionic theory obtained by $\Z_2^f$
gauging, which does not modify the local dynamics.

In the flux-attached theory \eqref{loop-FA}, the equations of motion
allow non-trivial vacuum fluctuations, as described earlier, thus the
Wilson loop operators are renormalized, and it is not clear how to
associate them to solitonic excitations.  Fortunately, an alternative,
nonperturbative definition for solitonic correlators was given in
\cite{frohlich87,frohlich95} and employed in Ref. \cite{3dbosonization2024}
(see Section 3, Eq. (3.26) in particular). It reads \be
\label{np-corr}
\langle \Phi(x_1) \Phi (x_2) \rangle = \exp \{ -S[f(x_1)-f(x_2)]\}\,,
\ee
where the field $\Phi(x)$ is implicitly defined by creating a localized flux
at $x=x_1$. This can be realized by imposing boundary conditions
for the path integral, or, equivalently, 
by inserting a Euclidean monopole $f(x)$ of appropriate flux.
The correlator \eqref{np-corr} is thus found by evaluating the
classical action $S$ in presence of fluxes. 

This definition can also be used in the flux-attached theory, within
the semiclassical approximation. The
path-integral should be evaluated  using the reduced
action $S_f[a,0]$ in \eqref{f-loop-a}
for $\d a$ fluxes and  $S_f[\z,0]$ in \eqref{f-loop-z}
for $\d\z$ fluxes. This calculation has already
been done in Ref.\cite{3dbosonization2024} (cf. Appendix A),
leading to the power-law result
\be
\label{np-corr-2}
\langle \Phi(x_1)_{N_0} \Phi (x_2)_{N_0} \rangle =
\frac{1}{(x_1-x_2)^{2\D_\Phi}}\,.  \ee In this expression, the
conformal dimension is given by $\D_\Phi=\D_{N_0=M_0}$ in
\eqref{spectra}, by taking into account that $\Phi$ creates
both magnetic and electric fluxes. The result matches the earlier
derivation of the partition function \eqref{Z-f-loop}. 

The fermionic action $S_f[a,0]$ \eqref{f-loop-a}
also contains the Chern-Simons term which determines the monodromy
 between $\Phi$ fields in the correlator $\langle\Phi(x_1)\Phi(x_2)\rangle$.
This is computed by considering the parameterization
$z=x+iy=\exp(i\vf(t))$ realizing the monodromy loop of $x_1=(t,x,y) $
going around $x_2=(0,0,0)$ and evaluate the change in Chern-Simons action
due to $\vf(t)\in[0,2\pi]$.  This clearly gives the
expected phase $\th =2\pi N^2_0/k$ in \eqref{f-monod}
(which should be included in the
expression \eqref{np-corr-2}, actually).

%-4.3.3---------------------------------
\subsubsection{Fermionic particle-vortex duality}

The generalized particle-vortex duality \eqref{bosonic PV with k}
introduced earlier in bosonic loop model can be rewritten,
\be
\label{bose-PV}
S_b[A] \qquad\longleftrightarrow\qquad 
S_b[kc] +\frac{1}{2\pi}c\d A = \wt{S}_b[A]\,.
\ee
It involves the $S$ transformation \eqref{SL2Z generators},
extended to $k\neq 1$
by letting $A\to k c$ and the usual coupling to $A$.
In the fermionic case, we are going to consider the following map
\be
\label{fermi-PV}
S_f[A] \qquad\longleftrightarrow\qquad 
S_f[kc] +\frac{1}{4\pi}c\d A = \wt{S}_f[A]\,.
\ee
This generalizes  the standard fermionic duality,
in its local form \eqref{ff duality wrong} for simplicity, 
by setting $A\to k c$ again.
Note also that the anomaly-canceling terms in \eqref{ff duality wrong}
are not considered, because anomalies are 
explicit in effective action approaches.

Upon applying the transformation in the action $S[a,\z,A]$  \eqref{f-loop},
and integrating the field $c$, the same action is obtained,
but time-reversal transformed,
\be
\label{f-loop-TR}
\wt{S}_f[a,\z,A]=\frac{g_0}{4\pi} \int a D a
+\frac{i}{2\pi}\int k \z \d a +\frac{k}{2} \z\d\z-\z\d A +
\frac{1}{4k} A\d A\,.
\ee
This result is not surprising because the flux-attached theory is already
particle-vortex symmetric within the same Lagrangian formulation
(cf. \eqref{f-loop-a} and \eqref{f-loop-z}).
Note also that the interplay between time reversal
and particle-vortex symmetry was already observed for
the Dirac-boson duality in Section \ref{Abelian dual sec}:
there we found the complementary feature
that time reversal symmetry is mapped by flux attachment
into bosonic particle-vortex duality.

In conclusion, we have discussed the properties of the fermionic
loop model obtained by flux attachment. This theory is different from that
found by $\Z_2^f$ gauging, in particular for the spectrum.
Nonetheless, we have checked that it is selfdual under
the (generalized) fermionic particle-vortex maps.

%-5-----------------------------------------------
\section{Dualities combining $\Z_2^f$ gauging and
  flux attachment}
\label{Bosonization of FA sec}

In the previous Sections, we described two methods relating fermionic
and bosonic theories, the $\Z_2^f$ gauging \eqref{oldbosonization} and
flux attachment \eqref{fluxattachment}, and found that they are
fundamentally different.  An interesting consequence is that the two
maps can be combined to obtain new kinds of dualities.  We  start
from the boson-fermion duality \eqref{dirac = boson + FA} (the seed
duality), obtained by flux attachment.  Since this is a map between
two fermionic theories, we can gauge the $\Z_2^f$ fermion parity  on
both sides to obtain a new boson-boson IR duality. This can be
generalized: for every fermionic IR duality, it is possible to obtain
its bosonic dual using \eqref{oldbosonization}.

In the following, we illustrate this idea in detail and later extend
it to the corresponding flux attachment for the Majorana
fermion \cite{metlitski2017MajoranaDuality,
  AharonyBeniniHsinSeiberg2017CSSODualities,3ddualityreviewSenthilSon}.

%-5.1------------------------------------------
\subsection{$\Z_2^f$ gauging of Dirac-boson duality}
\label{bos of Dirac sec}
\subsubsection{Bosonization}

We start by recalling the flux attachment \eqref{dirac = boson + FA}
for convenience
\begin{equation}
\label{dirac = boson + FA2}
\Bar{\psi} i \cancel{D}_A \psi \qquad
\longleftrightarrow \qquad |D_a\phi|^2 + V(|\phi|)+ \frac{i}{4\pi} a\d a +
\frac{i}{2\pi} a \d A.
\end{equation} 
The rhs is fermionic because of the level one Chern-Simons action.
We want to obtain its bosonic dual by gauging its fermion parity $\Z_2^f$.
To this effect, it is enough to remember the comment made in
Section \ref{U(1) CS is spin}: in this theory, the fermion parity
is generated by the reduction mod 2 of the first Chern class
$\d a/2\pi$, according to \eqref{CS depends on spin}. 

We introduce the background field $s$ for $\Z_2^f$ as in
\eqref{invertible bosonization}, with the coupling $(\d a/2\pi)s$. Then,
 we regard $s$ as a familiar $\U(1)$
gauge field and we introduce a Lagrange multiplier $c$ that enforces
its $\Z_2$ nature, such that only $\Z_2$ configurations of $s$ are
actually summed over. The resulting theory is:
\begin{equation}
  \int |D_a\phi|^2 + V(|\phi|)+ \frac{i}{4\pi} a\d a +
  \frac{i}{2\pi}s\d a - \frac{2i}{2\pi}s\d c + \frac{i}{2\pi} a \d A\,,
\end{equation}
(the sign of the Lagrange multiplier is chosen for later
convenience). Now it is straightforward to integrate out $s$, which
yields $a=2c$. Substituting this back in the action and renaming the
dynamical gauge field $a$ yields
\begin{equation}
\label{bosonic dual of scalar FA}
\int |D_{2a}\phi|^2 + V(|\phi|)+ \frac{4i}{4\pi} a\d a +
\frac{2i}{2\pi} a \d A.
\end{equation}
This is the bosonic dual of the rhs of the duality \eqref{dirac =
  boson + FA2}. It is a complex scalar field of charge $q=2$ coupled to
$\U(1)_4$ CS (the covariant derivative with subscript $qa$ stands for
$    |D_{qa}\phi |^2 = (\d + qia)\phi^\dagger \wedge *(\d-qia)\phi$).

\begin{figure}
 \begin{center}
  \includegraphics[scale=0.16]{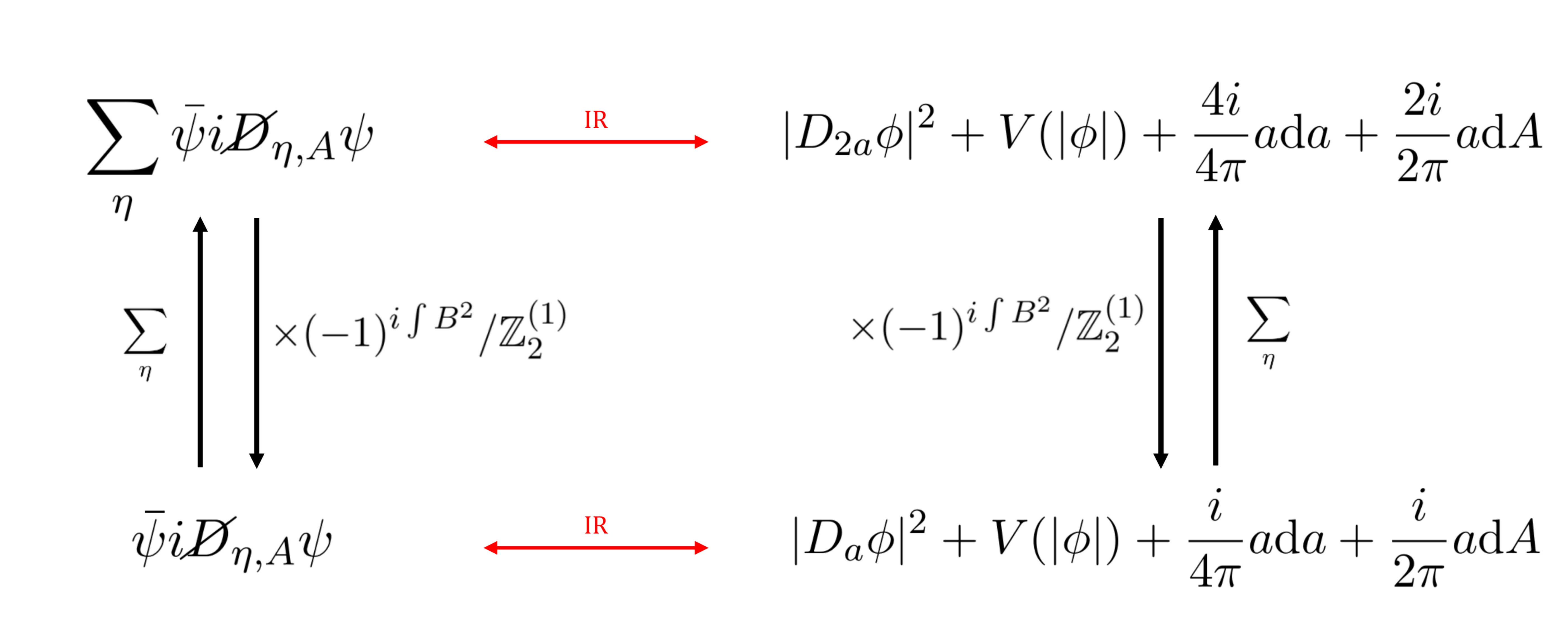}
 \end{center}
 \caption{The Dirac-boson flux attachment duality \eqref{dirac = boson
     + FA2} in the lower row. In the upper row, its bosonic dual
   obtained by using the bosonization map \eqref{oldbosonization}
   of $\Z_2^f$ gauging and its inverse \eqref{fermionization}.}
 \label{fig1}
\end{figure}

The same procedure can be done on the left hand side of the duality
\eqref{dirac = boson + FA2}. Summing over the spin structures 
gives a bosonic theory which should realize a bosonic IR
duality with \eqref{bosonic dual of scalar FA}.
All these maps form the diagram in Figure
\ref{fig1}. We conclude that the $\Z_2^f$ gauging 
of a Dirac fermion in 2+1d is IR dual to a charge-two complex scalar coupled
to $\U(1)_4$ CS theory,
\begin{equation}
\label{3d bosonization of dirac}
\sum_{\eta} \Bar{\psi} i \cancel{D}_{A,\eta} \psi \qquad
\longleftrightarrow \qquad |D_{2a}\phi|^2 + V(|\phi|)+ \frac{4i}{4\pi} a\d a +
\frac{2i}{2\pi} a \d A,
\end{equation} 
which can also be represented schematically
\begin{equation}
\label{3d bosonization of dirac 2}
\sum_{\eta} \text{Dirac fermion}( \eta) \qquad\longleftrightarrow\qquad
\phi \text{ (charge 2)} + \U(1)_4 \text{ CS} .
\end{equation} 
This realizes the bosonization map \eqref{oldbosonization} with
$B=0$. It is a new bosonic duality obtained by combining flux
attachment \eqref{dirac = boson + FA} and $\Z_2^f$ gauging.

%-5.1.2--------------------------------------
\subsubsection{Fermionization} 
It is instructive to run the inverse fermionization procedure starting from
\eqref{3d bosonization of dirac}. This shows the nature of the
bosonic dual symmetry $\Z_2^{(1)}$ with anomaly \eqref{3d bosonization
  anomaly} and serves as a check of the previous derivation.
In the next Section, this will also lead to a new
mixed anomaly in the more general case of manifolds with spin$_c$ structure.

The anomalous $\Z_2^{(1)}$ symmetry of \eqref{bosonic dual of scalar
  FA} is the subgroup of the  $\Z_4^{(1)}$ symmetry of
$\U(1)_4$ CS which survives after the introduction of charged $q=2$
matter \cite{gaiotto2015generalized}. This is a general fact. Consider
a complex scalar field of charge $q\in \Z$ coupled to $\U(1)_k$ CS,
\begin{equation}
\label{scalar charge q U(1)_k}
  S=\int_X  |D_{qa}\phi|^2 + V(|\phi|)+ \frac{ik}{4\pi} a\d a.
\end{equation}
The $\U(1)_k$ CS theory has a $\Z_k^{(1)}$ symmetry whose conserved
current is $J^{(2)} = * a$ (i.e. $\d a = 0$), but this conservation
law is broken in presence of charged matter, since the equations of
motion of \eqref{scalar charge q U(1)_k} are
\begin{equation}
    \d * J^{(2)} = \d a = \frac{2 \pi}{k}* J^{(1)},
\end{equation}
where $J^{(1)}$ is the $\U(1)$ matter current that is
coupled to $a$ (notice that for a scalar field it also contains a term
linear in $a$). However, the $\Z_k^{(1)}$ symmetry is not completely
broken in some cases. If we consider two Wilson lines $W(\gamma)$ and
$W(\gamma')$ such that $\p \Sigma = \gamma -\gamma'$, then
\begin{equation}
  W(\gamma) W^\dagger(\gamma') = e^{i \int_\Sigma a}=
  e^{\frac{2\pi i }{k} \int_\Sigma * J^{(1)}}=  e^{\frac{2\pi i }{k} q\Z},
  \qquad Q = \int_\Sigma * J^{(1)} = q\Z.
\end{equation} 
It follows that for $q$ multiple of $k$, the Wilson lines are still topological
and the $\Z_k^{(1)}$ symmetry is unbroken. More generally,
if $L = \gcd (k, q)$, then a subgroup $\Z_L^{(1)} \subset \Z_k^{(1)}$
is preserved. Indeed, if there is $m\in \Z$ such that $k = Lm$, then the line
$e^{im \oint a}$ is topological and
$(e^{im \oint a})^L = e^{i k \oint a} =1$.

Now, consider the bosonic theory \eqref{scalar charge q U(1)_k} with
a scalar field of charge $q=2$ and $\U(1)_{4}$ CS. The $\Z_{4}^{(1)}$
symmetry of $\U(1)_{4}$ is reduced to
$\Z_{\gcd (4,2)}^{(1)}= \Z_2^{(1)}$: this is the anomalous
$\Z_2^{(1)}$ symmetry due to the spin structure sum.
Following the by now standard fermionization procedure, we 
couple the theory \eqref{scalar charge q U(1)_k} to a $\Z_2$
background field $B$ for this symmetry. This can be done
in a manifestly gauge invariant way as follows
\begin{equation}
\label{bosonic dual of scalar FA with B}
S= \int_X |D_{2a}\phi|^2 + V(|\phi|)+
\int_Y \frac{4i}{4\pi} (f-B) \wedge (f-B), \qquad f = \d a\,,
\end{equation}
having used the 3+1d form \eqref{CS in 4d} for the Chern-Simons action.
This action \eqref{bosonic dual of scalar FA with B} involves the expression
\begin{equation}
    \frac{i}{\pi} \int_Y  B \wedge B \,,
\end{equation}
which cannot be written as a 2+1d term in $X$ nor it can be
though of as well-defined extension to $Y$ of a quantity in $X$.
Indeed, this is
precisely the anomaly for the $\Z_2^{(1)}$ symmetry for bosonization
\eqref{3d bosonization anomaly}.  Therefore, this $\Z_2^{(1)}$ symmetry
is correctly the dual symmetry of $\Z_2^f$.\footnote{ In Appendix
  \ref{app U4k to Uk}, we analyze in some detail the anomaly of the
  $\Z_k^{(1)}$ symmetry in $\U(1)_k$ CS for general $k$.}

Next, we make $B$ dynamical. This is a
$\Z_2$ field, so it does not affect the local dynamics. The new gauge
invariant field strength $f-B$ has semi-integer fluxes, so it is not
an $\U(1)$ curvature, but $2(f-B)$ is. We can thus introduce a new
$\U(1)$ connection $\Tilde{a}$ such that
$\Tilde{f}=\d \Tilde{a} = 2f-2B$. In terms of this variable, the
gauged theory is ($2 a \to \Tilde{a}$)
\begin{equation}
 S= \int_X | D_{\Tilde{a}} \phi |^2 + V(|\phi|) +
  \int_Y \frac{i}{4\pi} \Tilde{f} \wedge \Tilde{f},
\end{equation}
which is indeed the scalar field with flux attachment on the rhs of
\eqref{dirac = boson + FA2}. We thus checked the inverse map from the
bosonic \eqref{bosonic dual of scalar FA} to the fermionic
\eqref{dirac = boson + FA2} theories.

%-5.1.3----------------------------------------
\subsubsection{Spin$_c$ structure and mixed anomaly}
In the flux attachment \eqref{dirac = boson + FA2}, the electromagnetic field
$A$ has an important role: by keeping track of the $\U(1)$
symmetries in the two theories, it shows that \eqref{dirac = boson +
  FA2} is really a kind of particle-vortex duality. In the bosonic
duality we can read the coupling to $A$ from \eqref{bosonic dual of
  scalar FA}: the correct normalization has a factor two, which
follows after the $\Z_2^f$ gauging.
We can repeat the fermionization \eqref{bosonic dual of scalar FA with
  B} for $A \neq 0$, to obtain
\begin{equation}
S=  \int_X |D_{2a}\phi|^2+V(|\phi|)+
  \int_Y \frac{4i}{4\pi} (f-B) \wedge (f-B)+\frac{2i}{2\pi}(f-B)\wedge \d A.
\end{equation} 
There are now two terms that are apparently four-dimensional:
\begin{equation}
\label{spinc anomaly for dual scalar}
\int_Y \frac{i}{\pi} B \wedge B-\frac{2i}{2\pi} B\wedge\d A =i
\pi \int_Y  B \cup B+ B \cup \frac{2\d A}{2\pi}.
\end{equation}
In the last expression we switched back to the cohomological notation \eqref{3d Sq2B},
with $B\in H^2(Y,\Z_2)$ and the integer first Chern class $\d
A/2\pi$. The first term is the anomaly \eqref{3d bosonization
  anomaly} already discussed. The second term, involving $A$, is
actually independent from the four-dimensional extension, since it is
always $2\pi\Z$ on closed $Y$, so it is not a new anomaly.

Here there is an interesting generalization. What said is valid
for $A$ a standard $\U(1)$ connection. However, the
duality \eqref{dirac = boson + FA2} is actually defined for a more
general connection $A$, called spin$_c$ connection, which exists
for systems with spin-charge pairing
\cite{wittenwebofduality,seibergwitten2016gappedTI}. More precisely,
when excitations with half-integer spin have
odd charge and those with integer
spin have even charge (e.g. non-relativistic electronic systems and
QED), such a connection allows to define the fermionic theory
on non-spin manifolds with 
$[w_2]\neq 0$: non-trivial fluxes of $w_2$ are
compensated by those of $A$. Specifically, a spin$_c$ connection is
locally an $\U(1)$ connection with twisted flux
quantization\footnote{Note that the coupling
  $\int \frac{1}{4\pi}a \d a + \frac{1}{2\pi}a\d A$ in \eqref{dirac =
    boson + FA} is well-defined for $A$ a spin$_c$ connection
  \cite{seibergwitten2016gappedTI}.}
\begin{equation}
    \int_\S \frac{\d A}{2\pi} +\frac{1}{2}w_2 \in \Z,
\end{equation}
with $\S$ a two-cocycle (see also Appendix \ref{appBosonizationSpinc}
for its derivation). When $w_2\neq 0$, it follows that $\d A/2\pi$ has
half-integer fluxes, so $2\d A /2\pi$ could also assume odd integer
values when integrated on a closed manifold. After $\Z_2^f$ gauging,
we got the theory \eqref{bosonic dual of scalar FA} coupled to this
connection. In this case, the extra term involving $A$ in \eqref{spinc
  anomaly for dual scalar} is now dependent from the extension and it
is a mixed anomaly between the symmetries $\Z_2^{(1)}$ and $\U(1)$.
According to the duality \eqref{3d bosonization of dirac},
 this mixed anomaly should also be present in the bosonized
Dirac fermion coupled to a spin$_c$ connection. This is indeed the
correct anomaly expected from a general discussion regarding how
$\Z_2^f$ is related to the rest of the symmetry structure, as
explained in detail in Appendix \ref{appBosonizationSpinc} (see
\eqref{bosonization anomaly with spinc}).

Notice that in the bosonic theory \eqref{bosonic dual of scalar FA}
the effective connection probed by the boson is $2A$, which is a
standard $\U(1)$ connection with integer fluxes and it does not
require a spin$_c$ structure to be defined.

%-5.2-------------------------------------
\subsection{$\Z_2^f$ gauging of Majorana-boson duality}
We now consider the extension of the previous analysis 
to the bosonization of a Majorana fermion in three
dimensions. The goal is to obtain an analogue of \eqref{3d
  bosonization of dirac}. This is discussed also in
\cite{CordovaHsinSeiberg2018SON(n)CS}.

The bosonization of a Majorana fermion is obtained by a non-Abelian
extension of flux attachment
\cite{metlitski2017MajoranaDuality,AharonyBeniniHsinSeiberg2017CSSODualities}.
We can write it in a similar way to \eqref{dirac = boson + FA2}, as follows
\begin{equation}
\label{majorana = real boson + SO}
\Bar{\chi} i \cancel{D} \chi \quad \longleftrightarrow
\quad |D_a\phi|^2 + V(|\phi|)+ S_{\SO(N)_{-1}}[a] + N S_g.
\end{equation} 
On the lhs, there is a free
Majorana fermion $\chi$, coupled to gravity, namely the covariant derivative
$D_\mu$ involves the spin connection $\w_\mu$. On the rhs, a
real scalar $\phi$ in the vector representation of $\SO(N)$ is coupled
to a $\SO(N)$ gauge field $a$ and the non-Abelian Chern-Simons action is
introduced
\begin{equation}
\label{CS SO(N)}
S_{\SO(N)_k}[a;X]= \frac{ik}{8\pi} \int_X \Tr \left(a \d a +
  \frac{2}{3}a^3\right) = \frac{ik}{8\pi} \int_Y \Tr (f \wedge f),
\qquad \p Y = X.
\end{equation}
In this expression,
the trace is in the vector representation of $\SO(N)$, and $k\in \Z$.
Its value is taken to be $k=-1$. Similar to \eqref{CS in
  4d}, the form in $Y$ is actually independent from the
extension, and it is a more proper definition of the action 
when the gauge bundle is non-trivial.

The last term in \eqref{majorana
  = real boson + SO} is a gravitational background term needed to
match the phase diagram of the Majorana fermion on the left
\cite{metlitski2017MajoranaDuality},\footnote{We are using the sign
  convention of \cite{wittenwebofduality} for \eqref{Z SO(N)_1}. The
  duality \eqref{majorana = real boson + SO} is then fixed by
  requiring that the Majorana fermion realizes the trivial phase for
  $m>0$ and the topological superconductor $\SO(1)_1$,
  i.e. $e^{-S_g}$, for $m<0$. This amounts to regularize the
  fermion path integral as $Z=|Z| e^{-i\frac{\pi}{4}\eta}$. See also
  the comment for the Dirac case below \eqref{dirac = boson + FA}. The
  factor of two of difference with the Dirac case is because for the
  Majorana fermion we consider the Pfaffian instead of the determinant
  \cite{witten2016fermion}.}
\begin{equation}
\label{CS_g}
    S_g[X] = \frac{i}{192\pi} \int_Y \Tr (R \wedge R).
\end{equation}
Here, we have defined $S_g$ directly in terms of the extension to
$Y$ as in \eqref{CS SO(N)}, with $R$ the Riemann tensor on
$Y$. Interestingly, the $\SO(N)_1$ theories are invertible
topological field theories whose partition function is given precisely
in terms of $S_g$
\cite{wittenwebofduality,seibergwitten2016gappedTI},
\begin{equation}
\label{Z SO(N)_1}
    Z(\SO(N)_1) = e^{-NS_g}.
\end{equation}
It is thus possible to interpret  the rhs of
\eqref{majorana = real boson + SO} as a gauge theory of a scalar
interacting with an $\SO(N)$ gauge field tensored with a copy of
$\SO(N)_1$, which gives the correct gravitational background.

The duality \eqref{majorana = real boson + SO} was proposed in
\cite{metlitski2017MajoranaDuality,AharonyBeniniHsinSeiberg2017CSSODualities},
and it is a special case of a large class of
Chern-Simons plus matter dualities with orthogonal gauge groups.
A suitable quartic interaction in $V(|\phi|)$ is included to
tune the theory to the IR fixed point where
the duality holds. In addition, it is required that $N \geq
3$. First of all, it is a fermionic duality. On the rhs, the
Chern-Simons coupling introduces a spin structure dependence (for the
same argument of Section \ref{U(1) CS is spin}). Moreover, the
Majorana fermion is the monopole in the dual scalar theory, which
carries a $\Z_2$ charge identified with the fermion parity symmetry
$\Z_2^f$.\footnote{There is an additional $\Z_2$ symmetry on the rhs
  of \eqref{majorana = real boson + SO} given by O$(N)/\SO(N)$, where
  O$(N)$ is the global flavor symmetry before gauging. However, in
  \cite{metlitski2017MajoranaDuality} is argued that this symmetry is
  confined at the IR fixed point where the duality is conjectured to
  hold.} This symmetry is generated by $w_2(\SO)$, as explained around
\eqref{CS depends on spin}. 

As done for the Dirac fermion \eqref{dirac = boson + FA} in the
previous Section, we can sum over the spin structures on both sides of
\eqref{majorana = real boson + SO} to obtain a bosonic duality. The
left side gives a bosonization of the Majorana fermion. The rhs can be
treated as the $\Z_2^f$ gauging of a scalar field coupled to
$\SO(N)_{-1}$ CS with a torsion insertion given by $\SO(N)_1$
(this generalization is discussed in Appendix \ref{torsion app}).
The bosonization of the scalar theory with $\SO(N)_{-1}$ can be obtained
by gauging the symmetry generated by $w_2(\SO)$, according to
\eqref{CS depends on spin}. This effectively fixes $[w_2(\SO)]=0$: the
gauge bundle changes from $\SO(N)$ to $\spin (N)$. The scalar is then
in the $N$-dimensional representation of $\spin (N)$. The torsion insertion is treated as in \cite{HsinShao2020} (see
\eqref{Z2^f gauging with torsion} in Appendix \ref{torsion
  app}). 
  
All in all, it yields (remember $N\geq 3$)
\begin{equation}
\label{3d bosonization of Majorana}
\sum_{\eta} \text{Majorana fermion}(\eta) \qquad\longleftrightarrow\qquad
 \frac{(\phi \text{ ($N$-dim)} + \spin (N)_{-1})
    \times \spin (N)_1}{\Z_{2,\rm diag}^{(1)}}\,.
\end{equation} 
The quotient means gauging the diagonal $\Z_2^{(1)}$ symmetry given by
the anomaly-free subgroup of the two anomalous $\Z_2^{(1)}$ symmetries
(with anomaly \eqref{3d bosonization anomaly}) that arise after
$\Z_2^f$ gauging (the diagonal symmetry is generated by the bosonic
line given by the product of the two fermionic lines). As for the
Dirac case, \eqref{3d bosonization of Majorana} should be interpreted
as a purely bosonic IR duality between the two theories. The simplest
case is when $N=3$, where we can use the relations
$\SO(3)_k = \SU(2)_{2k}/\Z_2$ and $\spin(3)_k = \SU(2)_{2k}$
\cite{AharonyBeniniHsinSeiberg2017CSSODualities,
  CordovaHsinSeiberg2018SON(n)CS}. Therefore
the above duality reduces to
\begin{equation}
  \sum_{\eta}\text{Majorana fermion}(\eta) \qquad\longleftrightarrow\qquad
 \frac{(\phi \text{ (adj)} +
      \SU (2)_{-2}) \times \SU(2)_2}{\Z_{2,\text{diag}}^{(1)}}.
\end{equation} 
This is a new bosonic duality for the Majorana fermion analogous to
the Dirac case \eqref{3d bosonization of dirac 2}.

%-6--------------------------------------
\section{Conclusions}
In this work we analyzed the two bosonization approaches for relativistic
field theories in 2+1 dimensions: the sum over spin structures, or
$\Z_2^f$ gauging, and the flux attachment. We reviewed and summarized
their main features, including the extensions to non-spin and
non-orientable manifolds (in the Appendices).

We explicitly tested these bosonizations in the loop model, a
solvable conformal theory endowed with semiclassical dynamics.
This theory possesses a rich spectrum of electric and magnetic
solitons, which is invariant under particle-vortex duality. On the
basis of the detailed results in this example, we described the
differences between the two bosonizations approaches.
While the $\Z_2^f$ gauging modifies the
spectrum by introducing a selection rule between excitations (by gauge
invariance), the flux attachment strongly affects the dynamics beyond
the statistical transmutation. Furthermore. in the latter case the
resulting fermionic theory obeys particle-vortex selfduality, as expected,
also extended to fractional fluxes $k\neq 1$.

Next, we applied the two bosonizations together: we considered the
$\Z_2^f$ gauging on both sides of the flux attachment dualities for
Dirac and Majorana fermions (as shown in Figure \ref{fig1}). Both cases
gave new relations between three-dimensional bosonic theories. Given
the generality of the $\Z_2^f$ gauging, this method could be extended to
other known 2+1 dimensional fermionic dualities, and lead to
corresponding novel bosonic dualities.

Concerning possible developments of this work, we mention the extension
of $\Z_2^f$ gauging to 3+1 dimensions, which has already developed at
the level of topological field theories \cite{kapustinthorngren2017}
and lattice models \cite{kapustinBosLattice4d}. The challenge is to
discuss this approach in a toy example, generalizing the loop model,
which allows for a solvable dynamics. For example, it is conceivable
to complement the topological theory with the one-loop response term
$\int a_\mu \braket{J_\mu J_\nu} a_\nu$ which involve the universal
conformal correlator of two conserved $\U(1)$ currents in 3+1 dimensions.
Among other aspects, this extension is technically challenging due to
appearance of higher-form gauge fields and the associated extended
excitations and topological defects.

\bigskip

{\bf Acknowledgements} We would like to thank A. G. Abanov
for interesting discussions on the topics
of this work. This work has been partially supported by the
grants PRIN 2017 and PRIN 2022 provided by the Italian Ministery of
University and Research.

\appendix

%-A----------------------------------------
\section{$\Z_2^f$ gauging: details and extensions }
\label{appBosonization}
In this Appendix we collect some technical aspects and comments which
clarify the steps of the $\Z_2^f$ gauging procedure. In section
\ref{torsion app} and \ref{appBosonizationSpinc} we generalize this
bosonization to include torsion coefficients and twisted spin
structures
\cite{seibergwitten2016gappedTI,gaiottokapustinspinTQFT1,gaiottokapustinspinTQFT2,kapustinthorngren2017,kapustinFermionicBordism,thorngren2018thesisPHD,thorngren2020anomalies,HsinShao2020}.

\subsection{The $\Z_2^{(d-2)}$ anomaly and comparison
  with the literature}

The anomaly \eqref{Z_2^(d-2) anomaly} requires the knowledge of the
Steenrod squares and their relation with the Stiefel-Whitney classes
\cite{HatcherAT,kapustinthorngren2017}. The Steenrod squares are group
homomorphisms $Sq^p: H^k(X,\Z_2)\to H^{k+p}(X,\Z_2)$, which are
basically the generators of the algebra of operations between $\Z_2$
cohomology groups. They are defined axiomatically and can be expressed
using the higher cup products $\cup_i$
\cite{kapustinseiberg2014,kapustinthorngren2017}, i.e.
$Sq^p \alpha_k = \alpha_k \cup_{k-p} \alpha_k$ for
$[\alpha_k]\in H^k(X,\Z_2)$ (where
$\cup_i: H^p(X)\times H^q(X) \to H^{p+q-i}(X)$ and $\cup_0 = \cup$ is
the standard cup product). In particular,
$Sq^p \alpha_p = \alpha_p \cup \alpha_p$ and $Sq^p [\alpha_k] =0$ if
$p > k$. The Steenrod squares are related to the Stiefel-Whitney
classes by the Wu formula \cite{kapustinthorngren2017}:
$Sq^p [\alpha_{d-p}] = [v_p(TX)] \cup [\alpha_{d-p}]$, where
$[v_p(TX)]$ are polynomials in the Stiefel-Whitney classes called Wu
classes, whose expressions do not depend on the specific manifold
considered. The first four classes are $[v_0]=1$, $[v_1]=[w_1]$,
$[v_2]=[w_1]^2 + [w_2]$, $[v_3]= [w_1w_2]$, for example.

Using these operations, it is possible to express the bosonization
anomaly in $d+1$ dimensions in the form \eqref{Z_2^(d-2)
  anomaly}, which we rewrite here for convenience,
\begin{equation}
\label{anomaly fermionization app}
\mathcal{A} = i \pi \int_Y w_2 \cup B = i \pi \int_Y Sq^2B =
i \pi \int_Y B \cup_{d-3} B,
\end{equation}
where $B \in Z^{d-1}(Y,\Z_2)$ and $Y$ closed oriented. In the first
step of \eqref{anomaly fermionization app} we used the Wu formula
$Sq^2 [B] = [w_2] \cup [B]$ (since $[w_1]=0$ for $Y$ oriented), in the
second step we used the definition of $Sq^2B$ in terms of the higher
cup products. Notice that we are carefully taking track of the
cohomology classes, indicated by $[.]$, to which the Wu formula
applies. The reason why we can use the cohomology classes here is that
\eqref{anomaly fermionization app} is gauge invariant when $Y$ is
closed, so it is actually independent from the choice of
representatives. These three expressions for $\A$ define the same invertible
topological theory when $\p Y=0$, therefore they all describe the same
bosonic $\Z_2^{(d-2)}$-SPT phase (bosonic since $\A=0$ when $Y$ is
spin) and the same $d$-dimensional anomaly.\footnote{The same element
  of Hom$(\Omega_\SO^{d+1}(B^2\Z_2),\U(1))$ \cite{kapustinBordism}.}

We now compare the bosonization \eqref{fermionization} explained
in the main text, drawn from Ref. \cite{thorngren2018thesisPHD},
with the original formulation of Ref. \cite{gaiottokapustinspinTQFT1}. In the
latter work, bosonization was 
introduced as a way to describe spin TQFTs (with partition function
$Z_f$) starting from better understood
bosonic TQFTs (with $Z_b$). For a
large class of spin topological field theories, the partition
functions were related as follows
\begin{equation}
\label{GK bos}
\begin{split}
  Z_f[\eta] &= \sum_{B \in H^{d-1}(X,\Z_2)} Z_b[B](-1)^{\int_X \eta \cup B} 
\\
& = \sum_{B \in H^{d-1}(X,\Z_2)} \Tilde{Z}_b[B] \s[B] (-1)^{\int_X \eta \cup B}
=    \sum_{B \in H^{d-1}(X,\Z_2)} \Tilde{Z}_b[B] z[B,\eta],
\end{split}
\end{equation}
where 
\begin{equation}
\label{GK z}
    z[B,\eta] \equiv \s[B] (-1)^{\int_X \eta \cup B}.
\end{equation}
Every quantity depends on the topological space $X$,
defined by a triangulation of it. Notice that the 
expression in the first line of \eqref{GK bos} has the form
\eqref{fermionization} used in this work. In the second line, 
the original formulation in Ref. \cite{gaiottokapustinspinTQFT1}
is reported, where an
additional term  $\s[B]$ is extracted from the bosonic partition
function. This accounts for the
Grassmann--odd part of the fermionic path integral, in such a way that
the remaining $\Tilde{Z}_b$ is ``strictly'' bosonic.
It follows that all the
fermionic information of $Z_f$ is contained in the factor $z[B,\eta]$
\eqref{GK z}, which can be thought of as the partition function of a
``minimal'' spin TQFT.

Some properties define $z[B,\eta]$. Among them, its anomalous
variation is controlled by the Steenrod square of $B$ on $Y$, $Y=\p X$
($B$ is assumed to extend to $Y$),
\begin{equation}
\label{GK z anom}
    z[B+\delta B,\eta] = z[B,\eta] (-1)^{\int_Y \delta( Sq^2B)}.
\end{equation}
It follows that $\Tilde{Z}_b[B]$ is also anomalous with anomaly
$Sq^2B$: the product $\Tilde{Z}_b[B]z[B,\eta]$ is anomaly free and the
gauging procedure \eqref{GK bos} is well-defined. 

An explicit expression for \eqref{GK z} was given when $d\geq 3$
and the bulk $Y$ is orientable\footnote{This is the relevant case
  for this work. The $d=2$ case will be treated in the next Section,
  which is different because the anomaly \eqref{anomaly
    fermionization app} vanishes.} 
(always true in $d=3$),  as follows
\begin{equation}
\label{GK z explicit}
    z[B,\eta] = (-1)^{\int_X\eta \cup B +\int_Y Sq^2B+w_2\cup B}.
\end{equation}
This expression has indeed the anomaly \eqref{GK z anom}, because the
variation of the first and third terms cancel each other.
Note that \eqref{GK z explicit} is a purely $d$-dimensional term on $X$,
since it does not depend from the extension on $Y$.

In the following, we describe in a more detail the relation between
$z[B,\eta]$ and the phase $(-1)^{\int_X \eta \cup B}$, and thus the two
definitions of the bosonic partition function in \eqref{GK bos}.

The main observation is that the Wu formula evaluated on cocycles
has an unavoidable correction which is an exact form
\cite{gaiottokapustinspinTQFT1,thorngren2018thesisPHD}, namely
$Sq^2 B + w_2 \cup B = \d f(B)$.
This implies the following relation between the two forms of the anomaly
\eqref{anomaly fermionization app} on manifolds $Y$ with boundary $X$,
\begin{equation}
  \label{anom XY}
    i \pi \int_Y w_2 \cup B = i \pi \int_Y Sq^2B +
i \pi \int_X f(B).
\end{equation}
Using this relation, we can write \eqref{GK z explicit} as
\begin{equation}
    z[B,\eta] = (-1)^{\int_X\eta \cup B +f(B)}.
\end{equation}

Therefore, we have shown that the fermionic factors $(-1)^{\int_X \eta \cup B}$
and $z[B,\eta]$ in the two lines of \eqref{GK bos}
differ by the local counterterm $f(B)$ in $X$.
This means that they are characterized by equivalent forms of the same anomaly,
which is defined by the action \eqref{anomaly
  fermionization app} on $Y$ without boundary.
Nonetheless, they are different because the anomaly cancellations
involve the expression with boundary \eqref{anom XY}.
The local counterterms in $B$ in the
$d$-dimensional theory allow to pass from one form 
to the other.  In the first line of \eqref{GK bos},
the phase factor
$(-1)^{\int_X \eta \cup B}$ has the anomaly canceled by $\int_Y w_2 \cup B$. In
the second line, $z[B,\eta]$ has anomaly removed by $\int_Y Sq^2B$.

Summarizing, the relation between the two definitions of the bosonic
partition function involves the local counterterm $\int_Xf(B)$, 
\begin{equation}
    \Tilde{Z}_b[B] = Z_b[B](-1)^{\int_X f(B)}\,.
  \end{equation}
 
Given a bosonic theory $\T_b$ with anomaly \eqref{anomaly
  fermionization app}, the correct phase factor to use for obtaining the
fermionic theory, either $(-1)^{\int_X \eta \cup B}$ or $z[B,\eta]$,
depends on the form of the anomaly found after a
variation of $B$ in $\T_b$. Given the interpretation we gave of the
factor $\s[B]$ in \eqref{GK bos}, related to the Grassmann--odd part of
the functional integral, we expect that a bosonic theory written in
terms of bosonic variables only is naturally equipped with the $Sq^2B$
anomaly. This is what we found in Section \ref{loop-model Z2 gaug sec}
 for the loop model \eqref{k2-boundary-action} and in
Section \ref{bos of Dirac sec} for the scalar theory \eqref{bosonic dual of scalar FA with B}.
Notice in particular that we considered 4d spin manifolds $Y$,
for which the phase factor \eqref{GK z explicit} reduces to
\begin{equation}
    z[B,\eta] = (-1)^{\int_Y Sq^2B}=(-1)^{\int_Y B\cup B}\,.
  \end{equation}
  This is the form used for fermionizations discussed in
  Section \ref{loop model sec} and \ref{Bosonization of FA sec}.

%-A.2-------------------------------------
\subsection{Gauging fermion parity}
\label{app gauge Z_2^f or spin str}
In this work, we always assumed an equivalence between summing over
the spin structures and gauging fermion parity. This is realized by
saying that $\eta$ is the gauge field for $\Z_2^f$. However, in
Section \ref{bosonizationSEC} we also mentioned that a slightly
different interpretation is possible, which considers the gauge field
for $\Z_2^f$ as an element $s \in H^1(X,\Z_2)$, a standard choice for
an internal $\Z_2$ symmetry, that couples to the spin structure
dependence by shifting, $\eta \to \eta + s$. Here we comment on the
relation between these two interpretations.

Since any spin structure $\eta$ can be reached from a reference one by
$\eta_0 \to \eta_0 + s$, summing over $\eta$ or $s$ in $Z_f[\eta]$ gives
the same result, as showed in \eqref{oldbosonization}. This
is almost correct, but there is one subtlety. In general,
gauging a zero-form symmetry $A$ of a theory in $d$ dimensions $\T$
produces a gauged theory $\T / A$ with a dual $(d-2)$-form symmetry
$\Hat{A}^{(d-2)} \cong A^{(d-2)}$.\footnote{$\Hat{A}=$ Hom$(A, \R/\Z)$
  the Pontryagin dual group of $A$. For finite Abelian groups,
  $\Hat{A} \cong A$.} Gauging $\Hat{A}^{(d-2)}$ gives back the
original theory $\T$ \cite{bhardwajTachikawa2018,3dbosonization2024}:
this dual symmetry is not anomalous. On the contrary, as seen, gauging
the spin structures gives an anomalous $\Z_2^{(d-2)}$ symmetry with
anomaly \eqref{Z_2^(d-2) anomaly}. The difference between the two
gauging procedures is due to $\d \eta = w_2 \neq 0$, while $\d s = 0$.

This difference can be seen by turning on the background gauge fields,
as follows
\begin{equation}
\label{Z_2^f gauging}
\sum_{s\in H^2(X,\Z_2)} Z_f[\eta_0 +s] (-1)^{\int_X s \cup B} =
(-1)^{\int_X \eta_0 \cup B} \sum_\eta Z_f[\eta] (-1)^{\int_X \eta \cup B}=
(-1)^{\int_X \eta_0 \cup B} Z_b[B].
\end{equation}
Comparing with \eqref{invertible bosonization} shows that the
difference is in the coupling with the $B$ field: here the current is
$s$ (conserved $\d s=0$), not $\eta$ (not conserved $\d \eta \neq
0$). As a result, the gauging procedure \eqref{Z_2^f gauging} does not
yield a bosonic theory, but another fermionic theory, due to the
remnant $\eta_0$ dependence after
gauging. However, this fermionic theory is rather trivial, being just
$Z_b[B]$ in \eqref{invertible bosonization}, the result of the sum over
spin structures, times the extra phase in $\eta_0$.

We conclude that the two approaches are basically equivalent. The
difference between summing over $\eta$ or gauging $s$ boils down to an
extra factor $(-1)^{\int \eta \cup B}$ in the final theory
that can be removed by hand. At the end of the day, one can
always extract the same $Z_b[B]$ from $Z_f[\eta]$ in both approaches.

We can use this result to give an interpretation of the coupling
$(-1)^{\int \eta \cup B}$. Consider the final term of \eqref{Z_2^f
  gauging},
\begin{equation}
\label{Z_b times 1_spin}
   (-1)^{\int_X \eta \cup B} Z_b[B].
\end{equation}
This is $Z_b[B]$ tensored with a fermionic theory whose partition
function is $(-1)^{\int \eta \cup B}$. This fermionic theory is an
invertible topological theory with trivial partition function on every spin
manifold when $B=0$, but the dependence on the spin structure $\eta$
allows to write a coupling with a $(d-1)$-form field $B$. We thus
propose that $(-1)^{\int \eta \cup B}$ could be a representative of
the trivial spin TQFT in dimension $d$. Then \eqref{Z_b times 1_spin}
is $Z_b[B]$ tensored with this spin TQFT, both being
coupled to the same background field $B$. 

We can try to relate this discussion to explicit examples in low
dimensions. In $d=2$, it is common in the literature to define the
bosonization map using the gauging \eqref{Z_2^f gauging},
with the following expression
\cite{tongArf,backfiringbosonization2024,shaoTASI2023}
\begin{equation}
\label{2d bosonization Arf}
Z_b[B] = (-1)^{\arf(\eta+B)+\arf(\eta)}
\sum_s Z_f[\eta +s](-1)^{\int s\cup B},
\end{equation}
where the Arf invariant is the mod 2 index of the Dirac operator
\cite{tongArf}. The prefactor is needed to ensure that $Z_b$ is indeed
bosonic. The inverse fermionization procedure \eqref{fermionization}
is therefore\footnote{Notice that the phase factor given by the Arf
  invariant is the spin TQFT $z[B,\eta]$ \eqref{GK z} of
  \cite{gaiottokapustinspinTQFT1} in the two-dimensional case.}
\begin{equation}
\label{2d fermionization Arf}
Z_f[\eta+s] = \sum_B Z_b[B] (-1)^{\arf(\eta+B)+\arf(\eta)}
(-1)^{\int B\cup s}.
\end{equation}
By comparing \eqref{2d bosonization Arf} with
\eqref{Z_2^f gauging} (and \eqref{2d fermionization Arf} with
\eqref{fermionization}), we can identify
\begin{equation}
    (-1)^{\int \eta \cup B} \sim (-1)^{\arf(\eta+B)+\arf(\eta)}. 
\end{equation}
This identification is up to terms in $B$ that do no depend from
$\eta$ and it is the direct extension of the discussion of the
previous Section to the two-dimensional case. The rhs is indeed the
trivial spin TQFT when $B=0$. Moreover, this also shows from another
point of view why there is no anomaly \eqref{Z_2^(d-2) anomaly} in
$2d$. The bosonic partition function \eqref{2d bosonization Arf} is
invariant under $B\to B+\d \l$,
\begin{equation}
  Z_b[B+\d \l] = (-1)^{\arf(\eta+B+ \d \l)+\arf(\eta)}
  \sum_s Z_f[\eta +s](-1)^{\int s\cup B} (-1)^{ \int s\cup \d \l} = Z_b[B].
\end{equation}
Indeed, the variation of $\int s\cup \d\l$ vanishes after
integration by parts. Crucially, the first term with the $\arf$
invariant is also invariant given that $\eta \sim \eta +\d \l$ (they define
the same spin structure), so the gauge parameter can be reabsorbed in
$\eta$. Notice that this is a peculiarity of two dimensions, because
$B$ is a one form: for this reason, $\d \l$ can be combined with
$\eta$.

For $d=3$, in Ref.\cite{HsinShao2020} it is argued that gauging $\Z_2^f$ as
in \eqref{Z_2^f gauging} gives the bosonic theory tensored with
SO$(0)_1$ Chern-Simons theory. $\SO(0)_1$ is indeed the trivial spin
TQFT in three-dimensions \cite{wittenwebofduality,
  CordovaHsinSeiberg2018SON(n)CS}, consistent with the interpretation
of \eqref{Z_b times 1_spin}.

To conclude this discussion, we notice that the equivalence of the sum
over spin structures and the $\Z_2^f$ gauging using $s\in H^1(X,\Z_2)$
holds when there are no other anomalies. If there are, they
could lead to different obstructions for the two procedures, as
shown in detail in $2d$ in \cite{backfiringbosonization2024}. In three
dimensions there are no other anomalies, but the torsion elements
admissible in the gauging procedures are different
\cite{HsinShao2020}. To avoid these complications, we really just
define the background gauge field for $\Z_2^f$ (which is not an usual
internal symmetry, since it displays spacetime features) to be the
spin structure $\eta$, following a conjecture of
\cite{kapustinFermionicBordism}. With this convention, $\Z_2^f$
gauging is by definition the sum over the spin structures.

%-A.3-------------------------------------
\subsection{Torsion coefficients}
\label{torsion app}
In this Section we describe some extensions of the $\Z_2^f$ bosonization
\eqref{invertible bosonization}. Much like for standard gauging
procedures, torsion coefficients can be inserted before doing the sum
(or the integral). These amount to extending the theory with
additional SPT phases, which are gauge
invariant by themselves. For example, in two dimensions,
the possible torsion coefficient is
the Arf invariant, i.e. the Kitaev chain SPT
\cite{tongArf,HsinShao2020,backfiringbosonization2024,
  thorngren2020anomalies}. In three dimensions, instead, the possible torsion
terms are partition functions of 
$\SO(n)_1$ Chern-Simons theories \cite{seibergwitten2016gappedTI},
which correspond to invertible TQFTs (for example, those of the
three-dimensional SPT phase called $p+ip$ superconductor)
\cite{kapustinFermionicBordism}. As a consequence, the alternative
bosonizations lead to bosonic theories related to each other by
stacking $\spin (n)_1$ CS theories and discrete gaugings
\cite{HsinShao2020}. We review below this three-dimensional case.

The $\Z_2^f$ gauging \eqref{invertible bosonization} with the
insertion of a torsion coefficient reads
\begin{equation}
\label{16 ways of Bos}
    Z_b^n[B] = \sum_{\eta} Z_f[\eta] Z(\SO(n)_1)[\eta](-1)^{\int_X\eta \cup B}.
\end{equation}
The SO$(n)_1$ CS theories are invertible TQFTs: they are
fermionic SPT phases whose Wilson lines are $\{1, \psi\}$, with $1$
the identity line and $\psi$ a neutral fermionic line. They are linear
in the gauge group rank ($\SO(n)_1 \times \SO(m)_1 = \SO(n+m)_1$) and
their partition function is just a phase
\cite{tongwebofduality,seibergwitten2016gappedTI}
\begin{equation}
    Z(\SO(n)_1) = e^{- n S_g},
\end{equation}
where $S_g$ is the gravitational Chern--Simons term
\eqref{CS_g}. $\SO(0)_1$ is the trivial spin TQFT.

According to \eqref{16 ways of Bos}, several bosonic theories $Z_b^n[B]$ can be obtained from a fermionic one. However, they are related to each other
\cite{HsinShao2020}. Let us rewrite equation \eqref{16 ways of Bos} in
the equivalent form
\begin{equation}
\begin{split}
  Z_b^n[B] &=\sum_{B'\in H^2(X,\Z_2)}
  \sum_{\eta,\eta'} Z_f[\eta] Z(\SO(n)_1)[\eta']
  (-1)^{\int_X (\eta-\eta')\cup B'} (-1)^{\int_X\eta \cup B} =
  \\
    &= \sum_{B'\in H^2(X,\Z_2)} Z_b^0[B'+B] Z(\spin (n)_1)[B'].
\end{split}
\end{equation}
The sums over $\eta$ and $\eta'$ separately give the bosonization of
$Z_f$ and $\SO(n)_1$ (which is Spin(n)$_1$ CS theory \cite{CordovaHsinSeiberg2018SON(n)CS,seibergwitten2016gappedTI}), with non-zero
$B$ field. Each of these two bosonic theories has an anomalous
$\Z_2^{(1)}$ symmetry dual to the sum over spin structures. Then, the
sum over $B'$ is equivalent to gauge the non-anomalous diagonal $\Z_{2,diag}^{(1)}$ symmetry generated by the
bosonic line given by the product of the fermionic line of $Z_b^0$
and the fermionic line of $\spin(n)_1$ (indeed
originally $B'$ couples to $\eta -\eta'=s$, and $\d s =0$). So:
\begin{equation}
\label{Z2^f gauging with torsion}
\mathcal{T}_b^n  = \frac{\mathcal{T}_b^0 \times \text{Spin}(n)_1}
{\Z_{2,diag}^{(1)}}.
\end{equation}
Here $\T_b^n$ denotes the bosonic theory with partition function
$Z_b^n$ in \eqref{16 ways of Bos}. $\T_b^0$ corresponds to no torsion
insertions in the sum over spin structures, i.e. \eqref{invertible
  bosonization}. 

Inverting the bosonization with torsion \eqref{16 ways of Bos} as in
\eqref{fermionization}, one gets the `twisted' fermionic theory
\begin{equation}
    Z_f[\eta] Z(\SO(n)_1)[\eta] = \sum_{B} Z_b^n[B](-1)^{\int_X\eta \cup B},
\end{equation}
which is the original $Z_f[\eta]$ times an additional
gravitational response.

%-A.4-----------------------------
\subsection{Mixed anomaly and twisted spin structures}
\label{appBosonizationSpinc}
In the discussion on the bosonization \eqref{invertible bosonization}
in the main text, we focused solely on the spin structure dependence
and its intimately associated fermion parity symmetry
$\Z_2^f$. However, other symmetries could play an indirect role
in the bosonization map, mostly in terms of obstructions, given by
mixed anomalies, and the strictly related symmetry structures that
they form with $\Z_2^f$. In this Appendix we give a description of
such cases.

\subsubsection{Spin$_G$ structures}

This discussion is akin to usual characterizations of anomalies and
symmetries if we focus on the fermion parity symmetry $\Z_2^f$. This
is an intrinsic symmetry of every fermionic theory, generated by
$(-1)^F$, where $F$ is the number of fermions mod 2. This symmetry
extends the local Lorentz SO$(d)$ symmetry to Spin$(d)$,
necessary to define fermions. Mathematically, this implies a group
central extension, as follows,
\begin{equation}
    1 \to \Z_2^f \to \spin (d) \to \SO(d) \to 1,
\end{equation}
which means that Spin$(d)/\Z_2^f=$ SO$(d)$. As discussed, the
existence of a spin structure $\eta$, with $w_2 = \d\eta$, allows to
lift SO$(d)$ transition functions to $\spin (d)$ transition functions
consistently.

More generally, there can be other symmetries, besides Lorentz
symmetry, under which fermions transform. We call collectively the
`bosonic' symmetry group $G_b$. It is bosonic in the sense that it
acts at most on fermion bilinears (bosons), in some representation
$\rho$, and we want to understand its action on fermions. The
transition functions of the manifold are now valued in
$\SO(d) \times G_b$, which is this group that should be
extended by $\Z_2^f$ in order to define fermions, as follows,
\begin{equation}
\label{fermionic extension}
    1 \to \Z_2^f \to G_f \to \SO(d)\times G_b \to 1.
\end{equation} 
At the level of spin structures, instead of considering just the
tangent bundle $TX$, we are looking for a spin structure on the bundle
$TX \oplus T_\rho G_b$ \cite{kapustinFermionicBordism} (which has
indeed the structure group $\SO(d)\times G_b$), where $T_\rho G_f$ is
the vector bundle associated to the principal $G_b$ bundle in the
representation $\rho$. This generalized spin structure is sometimes called a
`twisted' spin structure \cite{thorngren2020anomalies} or
$\spin_{G_b}$ structure \cite{avisisham1980spinG}. It can be defined if
$w_2(TX \oplus T_\rho G_b)$ is exact (the argument is the same). It is
thus possible to define fermions even on non-spin manifolds if they
are charged and there exists a suitable twisted spin structure.

In the bosonization \eqref{invertible bosonization}, we map a
fermionic theory to a bosonic one by gauging $\Z_2^f$ (so we are
assuming here no anomaly for $\Z_2^f$). This gives a bosonic theory
with a dual $\Z_2^{(d-2)}$ symmetry with an 't Hooft anomaly
\eqref{Z_2^(d-2) anomaly}. But if the extension \eqref{fermionic
  extension} is not trivial, then after gauging there is a mixed
anomaly between $\Z_2^{(d-2)}$ and $G_b$
\cite{tachikawaGaugeFiniteGroups}. Neglecting the spacetime part
$\SO(d)$ for now, the central extension \eqref{fermionic extension} is
classified by an element $\a \in H^2(BG_b,\Z_2)$, where $BG_b$ is the
classifying space of $G_b$. Considering the gauge field $A$ for $G_b$ as a map from $X$ to $BG_b$ \cite{dijkgraafwitten}, the mixed
anomaly is
\begin{equation}
\label{extension mixed anomaly}
    \int_Y A^* \alpha \cup  B \,,
\end{equation}
(where $A^*$ is the pullback of the map $A$). This mixed anomaly
vanishes if $\alpha =0$, which means that the sequence involving the
$G_b$ part of \eqref{fermionic extension} splits, i.e
$G_f = G_b \times \Z_2^f$. It follows that the extension of $\Z_2^f$
acts only on the usual $\SO(d)$ part, i.e. it is possible to define
the spin structure by itself. If the manifold is not spin, the
symmetry has a non-trivial mixing with $\Z_2^f$ leading to the above
anomaly.

The anomaly \eqref{extension mixed anomaly} arises
in group extensions like \eqref{fermionic extension} since the gauge field
$s$ for the Abelian extension $\Z_2$ is not closed when $A \neq 0$,
but $\d s = A^* \a$. So, after gauging, the current ($s$ dynamical)
that generates $\Z_2^{(d-2)}$ is not conserved when $A \neq 0$ and the
coupling $s \cup B$ generates the anomaly \eqref{extension mixed
  anomaly}. Notice that this is the same argument that we used to
derive the bosonization anomaly \eqref{Z_2^(d-2) anomaly} given that
$\eta$ is the background field for $\Z_2^f$ and $\d \eta = w_2$. We
can conclude that the full anomaly when \eqref{fermionic extension} is
not trivial is
\begin{equation}
\label{full anomaly with twisted spin}
i\pi \int_Y Sq^2 B + A^* \alpha \cup B =
i\pi \int_Y (w_2(TY) + A^* \alpha) \cup B, 
\end{equation}
that can be traced to the fact that \eqref{invertible bosonization} is
now a sum over $\Tilde{\eta}$ with
\begin{equation}
\label{twisted spin structure}
    \d \Tilde{\eta} = w_2(TX) + A^* \alpha.
\end{equation}
$\Tilde{\eta}$ is the background field for $\Z_2^f$ and can be
interpreted as the twisted spin structure on $TX \oplus T_\rho G_b$,
i.e. $\d \Tilde{\eta} = w_2 (TX \oplus T_\rho G_b)$. The anomaly \eqref{full anomaly with twisted spin} can thus
be seen in two ways: from a more standard point view, it is the
anomaly arising after the gauging of $\Z_2^f$ given that the full
symmetry structure is \eqref{fermionic extension}; from a more
geometrical perspective, it follows from the sum over spin structures
\eqref{invertible bosonization}, where the spin structures are not
defined on $TX$ but on $TX \oplus T_\rho G_b$ and thus they satisfy
\eqref{twisted spin structure}.

Since we are considering orientable manifolds $[w_1(TX)]=0$, from the
general theory of characteristic classes \cite{nakahara}, the second
Stiefel-Whitney class on $TX \oplus T_\rho G_b$ is given
by\footnote{In this discussion we are also assuming that
  $[w_1(T_\rho G_b)]=0$, i.e. $TX \oplus T_\rho G_b$ is orientable
  (since we are looking for a spin structure).}
\begin{equation}
    [w_2(TX \oplus T_\rho G_b)] = [w_2(TX) + w_2(T_\rho G_b)].
\end{equation}
This allows to identify $w_2(T_\rho G_b)$ with $A^*\a$ in
\eqref{twisted spin structure} in this case.\footnote{Indeed,
  $\alpha \in H^2(BG_b, \Z_2)$ is the cohomology class of the
  classifying space $BG_b$, whose pullback under the map
  $A: X\to BG_b$ gives the second Stiefel-Whitney class for $G_b$
  bundles \cite{dijkgraafwitten}.} Notice that any spin manifold has
also a twisted spin structure: if $[w_2(TX)]=0$, it is enough to take
$T_\rho G_b$ as the trivial bundle $X \times_\rho G_b$
($[w_2(T_\rho G_b)]=0$). More generally, a twisted spin structure on
$X$ is possible when $[w_2(TX)] = [w_2(T_\rho G_b)]$. The strategy is
then to kill the non-trivial second Stiefel-Whitney class of $X$ with
the one of the gauge bundle. This implies a symmetry structure like
\eqref{fermionic extension} where $\Z_2^f$ is mixed with
$G_b$. Physically, it means that there is a relation between the spin
and internal symmetry quantum numbers, the charges.

%-A.4.2-----------------
\subsubsection{Spin$_c$ structures}
A particular important example for condensed matter systems (but also
QED) is the spin$_c$ structure
\cite{avisisham1980spinG,wittenwebofduality,seibergwitten2016gappedTI}. In
such systems, all excitations obey the spin-charge relation, since the fundamental degrees of freedom are electrons,
i.e. fermions in the fundamental representation (unit charge) of an
$\U(1)$ symmetry. The minimal representation $\rho$ for bilinears
(bosons) is thus that of charge two ($\rho =2$). The transition functions for
spinors\footnote{Spinors are sections of the product bundle
  $S \otimes T_1\U(1)$, where $S$ is the spinor bundle and $T_1\U(1)$
  the associated vector bundle of the principal $\U(1)$ bundle with
  unit charge.} are now a product $\l u$, where $\l \in \spin (d)$ and
$u\in \U(1)$. They are valued in the so-called Spin$_c$ group,
\begin{equation}
\label{Spinc group}
    \spin_c(d)= \frac{\spin(d)\times \U(1)}{\Z_2},
\end{equation}
where the $\Z_2$ quotient is $(\l,u)\sim (-\l,-u)$ (they give the same
transition function $\l u$).

In our language developed before, the relevant fermionic extension
\eqref{fermionic extension} to consider is
\begin{equation}
\label{spinc extension}
    1 \to \Z_2^f \to \spin_c (d) \to \SO(d)\times \U(1) \to 1.
\end{equation}
So $\spin_c (d) / \Z_2= \SO(d)\times \U(1)$ (instead of
$\spin (d) / \Z_2 = \SO(d)$). This indeed follows from the form of the
$\spin_c(d)$ group \eqref{Spinc group}: the first inclusion
$\Z_2^f \xhookrightarrow{} \spin_c(d)$ is ${\pm1}\to(\pm 1, 1)$, while
the projection map $\spin_c(d) \to \SO(d)\times \U(1)$ is
$(\l, u)\to (\pi(\l),u^2)$, where $\pi: \spin(d)\to \SO(d)$. These are
well-defined on the equivalence classes by the $\Z_2$ quotient of
\eqref{Spinc group} (using $\pi(\l) = \pi(-\l)$) and indeed the kernel
of the projection map is the image of the first inclusion, giving
\eqref{spinc extension}.

A spin$_c$ structure is a spin structure on $TX \oplus T_2 \U(1)$,
which exists when $w_2(TX \oplus T_2 \U(1)) \equiv w_2^c = \d
\eta^c$. Given that $w_2(\U(1))$ is the reduction mod 2 of the usual
$\U(1)$ first Chern class, we get\footnote{The obstruction to define a spin$_c$ structure on a manifold
  $X$ is given by the third integral Stiefel-Whitney class
  $[W_3(X)] \in H^3(X,\Z)$ \cite{avisisham1980spinG}. A
choice of spin$_c$ structure is an $\eta^c$ such that
$W_3= \d \eta^c$ and the number of inequivalent spin$_c$ structures on a
manifold $X$ is thus $H^2(X,\Z)$ (they are an affine space over
$H^2(X,\Z)$). In our language, $\eta^c$ is also given by \eqref{spinc
  structure}. It is possible to show that \eqref{spinc structure} gives indeed the same amount of inequivalent spin$_c$ structures.}
\begin{equation}
\label{spinc structure}
w_2^c = \d \eta^c = w_2(TX) + \left[\frac{\d A_2}{2\pi} \; \text{mod 2}\right]
= w_2(TX) + \left[\frac{2\d A}{2\pi} \; \text{mod 2}\right].
\end{equation}
In the last expression, we used the fact that the connection $A_2$ for
fermion bilinears on $T_2\U(1)$ is twice the one acting on the
fermions of charge one, so $A_2= 2A$.\footnote{Notice that if we
  normalize to one the charge of the bosons, the spinors are then in a
  projective half-integer charge representation of $\U(1)$. This is
  though a regular linear representation of the $\spin_c$ group
  \eqref{Spinc group} because of $\Z_2$ quotient.} The spin$_c$
structure is thus a combination of $w_2$ and $A$. If we divide the
above expression by two and integrate over a two-cycle $\S$, we obtain
the modified flux quantization found in the literature
\cite{wittenwebofduality,seibergwitten2016gappedTI,3ddualityreviewSenthilSon},
\begin{equation}
\label{spinc connection}
    \int_\S \frac{1}{2} w_2(TX) + \frac{\d A}{2\pi} = 0 \; \text{mod 1}.
\end{equation}
The gauge field $A$ is called a spin$_c$ connection, that can have
half-integer fluxes. In summary, it is possible to define consistently
charged-one fermions even on non-spin manifolds. At the level of
transition functions valued in \eqref{Spinc group}, this means that
$\spin(d)$ and $\U(1)$ transition functions are not good by
themselves, but their product is.

The anomaly \eqref{full anomaly with twisted spin} for the case of a
spin$_c$ structure is
\begin{equation}
\label{bosonization anomaly with spinc}
i\pi \int_Y Sq^2 B + \frac{2 \d A}{2\pi} \cup B =
i\pi \int_Y  \left(w_2(TY) + \frac{2 \d A}{2\pi}\right) \cup B. 
\end{equation}
This is the anomaly found in \eqref{spinc anomaly for dual scalar}. As remarked there, the extra piece in $A$ is
trivial if $A$ is a standard $\U(1)$ connection (and \eqref{spinc
  extension} on $\U(1)$ splits). On the other hand, when $A$ is a
spin$_c$ connection with half-integer fluxes, then its contribute is
an extra mixed anomaly term between $\Z_2^{(d-2)}$ and $\U(1)$.

In three-dimensions every orientable manifold is spin, so the
distinction between spin and spin$_c$ structures is not particularly
relevant. The spin$_c$ structure can be considered as a book-keeping
device to keep track of the underlying electronic nature of the
microscopic system considered \cite{seibergwitten2016gappedTI}. However, there are subtle effects
even in such cases, given that anomalies are characterized by inflow
from a four-dimensional bulk which can be non-spin but spin$_c$. If
we consider a three-dimensional system with a spin$_c$ structure, the
four-dimensional bulk will also be required to have a spin$_c$
structure, thus the extension of $A$ in \eqref{bosonization
  anomaly with spinc} is to a four-dimensional spin$_c$ connection. In
this case, the mixed anomaly term in \eqref{bosonization anomaly with
  spinc} is non-trivial. 

%-A.4.3----------------------------------------
\subsubsection{Non-orientable case: Pin$_G^{\pm}$ structures and time
  reversal symmetry}
We have considered so far orientable manifolds. However, theories with time-reversal symmetry $\cal T$ can be defined also on
non-orientable manifolds (the `background gauge field' for $\T$
\cite{witten2016parityanomalyUnoriented}). They can be described by extending the previous analysis.

There are two
possible ways to define fermions on non-orientable manifolds, which
correspond to the two possible double coverings of O$(d)$: one is
called Pin$^+$ structure (with transition functions in Pin$^+(d)$) and
it applies when $\T^2 = (-1)^F$, which is the most relevant case for
relativistic fermions; the other case is a Pin$^-$ structure, when
$\T^2=1$ \cite{witten2016fermion}. The obstruction for Pin$^+$
structures is still $w_2(TX)$ (but $w_1(TX)\neq 0$), while for Pin$^-$
is $w_2(TX) + w_1^2(TX)$
\cite{kapustinFermionicBordism}. Consequently, a Pin$^+$ structure is
still defined by $\eta^+$ such that $\d \eta^+ =w_2$, while a Pin$^-$
structure is $\eta^-$ such that $\d \eta^- = w_2 + w_1^2$. If we apply
the bosonization procedure \eqref{invertible bosonization} as a sum
over Pin structures, taking into account the above definition of
$\eta^{\pm}$ and also the non-trivial extension \eqref{fermionic
  extension} (which corresponds to Pin$^{\pm}_G$ structures in this
case), the dual bosonic theory has the following anomalies for the
arising dual symmetry $\Z_2^{(d-2)}$ (extending \eqref{Z_2^(d-2) anomaly}):
\begin{equation}
\label{bosonization anomaly with TR}
    \begin{split}
      &\text{Pin$^+$: } i \pi \int_Y (w_2(TY) + A^* \alpha) \cup B =
      i\pi \int_Y Sq^2 B +  w_1^2(TY) \cup B+ A^* \alpha \cup B; \\
      &\text{Pin$^-$: } i\pi \int_Y (w_2(TY) + w_1^2(TY) + A^* \alpha)
      \cup B = i\pi \int_Y Sq^2 B + A^* \alpha \cup B.
    \end{split}
\end{equation}
From this perspective, the role of time reversal is not encoded in
$G_b$ in the extension \eqref{fermionic extension}, but in the
spacetime term $\SO(d)$, which becomes O$(d)$ (indeed
Pin$^{\pm}/\Z_2 =$ O$(d)$ \cite{witten2016fermion}), but it is also
possible to see the Pin$^{\pm}$ structure as a twisted spin structure
itself \cite{thorngren2020anomalies}. In the Pin$^+$ structure case
there appears to be a mixed anomaly between time reversal and the
$\Z_2^{(d-2)}$ symmetry, the term $w_1^2 \cup B$, which is instead
absent in the Pin$^-$ case \cite{bhardwaj2017unorientedTFT3d}. This
mixed anomaly can be traced back to the fact that in the Pin$^+$ case
the bosonic time reversal symmetry $\T$ (which is a $\Z_2$ symmetry)
is non-trivially extended by fermion parity, given that
$\T_f^2 = (-1)^F$ (i.e., in the fermionic system, time reversal is a
$\Z_4^T$ symmetry given by the extension
$\Z_2^f \to \Z_4^T \to \Z_2^T$). Much like \eqref{fermionic extension},
there is then a mixed anomaly. In the main text, we found this mixed
anomaly in the relation \eqref{T-square}.

Notice that in the three dimensional case, the anomalies
\eqref{bosonization anomaly with TR} agree with the ones proposed in
\cite{bhardwaj2017unorientedTFT3d} when $A=0$ (formulas (3.10) and
(5.2) there).

\section{$\U(1)_{4k}$ Chern--Simons and
  its gauging to $\U(1)_k$}
\label{app U4k to Uk}
In this Appendix we show in some detail the relation between non-spin
$\U(1)_{4k} = \spin (2)_k$ and spin $\U(1)_k = \SO(2)_k$ CS theories
(which is a particular case of the relation between $\spin (n)_k$ and
$\SO(n)_k$ \cite{CordovaHsinSeiberg2018SON(n)CS}).

$\U(1)_k$ CS theory \eqref{CS in 4d}
has a $\Z_k^{(1)}$ symmetry generated by the
Wilson loops of $a$. At the level of the action, this can be seen by
the invariance under the shift $a \ra a+ \lambda$, with
$\d \lambda =0$. We can add a background field for this symmetry, $B$,
with the background gauge transformation
\begin{equation}
    B \ra B + \d \lambda, \qquad a \ra a+\lambda.
\end{equation}
The gauge invariant object is $f-B$, so it is clear how to couple $B$
in the action \eqref{CS in 4d}:
\begin{equation}
\label{CS gauged in 4d with anomaly}
S = \int_Y \frac{ik}{4\pi} (f-B) \wedge (f-B) =
\int_X \frac{ik}{4\pi} a\d a - \frac{ik}{2\pi}aB +\int_Y \frac{ik}{4\pi} B\wedge B.
\end{equation}
The whole action is gauge invariant. The first two terms are clearly
local on $X$ and gauge invariance under $\U(1)_a$
($a \ra a +\d\alpha$) imposes that $B$ is a $\Z_k$ gauge field
($\d B=0$ and $\oint B = 2\pi /k\Z$ for invariance under large gauge
transformations). These two terms alone, however, are not gauge
invariant for $\Z_k^{(1)}$ and the extra term $B^2$ on $Y$ is needed,
which is an anomaly if it depends on the extension to $Y$. Let us
check this. If $Y$ and $Y'$ are two different extensions, with
$Z =Y-Y'$, $\p Z =0$,
\begin{equation}
\label{'tHooft anomaly CS}
\A = S -S' = \int_Z \frac{ik}{4\pi} B \wedge B =
\frac{i\pi}{k} \int_Z \Tilde{B} \cup \Tilde{B} =
\frac{i\pi}{k} n, \qquad n=0,1,...,k-1.
\end{equation}
Here $B = 2\pi/k \Tilde{B}$, with $ \Tilde{B}$ the discrete gauge
field. So in general $S \neq S'$ and the action depends from the
extension. This shows that the $\Z_k^{(1)}$ symmetry has an 't Hooft
anomaly and cannot be gauged.

Consider the case with $k=2m$ even. There is always a $\Z_2^{(1)}$
symmetry, given that $\Z_2 \subseteq \Z_{2m}$, generated by the Wilson
line $e^{im \oint a}$. The spin of this quasiparticle is
\begin{equation}
\label{spin middle line}
    \spin (e^{im\oint a}) = \frac{m^2}{4m} =\frac{m}{4} \; \text{mod} \; 1 
    \begin{cases}
      \frac{1}{2} \; \text{if } m =  2 \; \text{mod } 4 \quad
      &\text{fermion;} \\
        0 \; \text{if } m =  0 \; \text{mod } 4 \quad &\text{boson;}\\
        \frac{m}{4} \; \text{if } m =  \text{odd} \quad &\text{anyon.}
    \end{cases}
\end{equation} 
Combine this with the anomaly \eqref{'tHooft anomaly CS}: if we gauge
a subgroup $\Z_2 \subseteq \Z_{2m}$ of the one-form symmetry (i.e. $B$
is a $\Z_2$ gauge field) the anomaly is\footnote{Notice that the
  anomaly for a $\Z_k^{(1)}$ symmetry generated by a line with spin
  $s$ is \cite{CordovaBeniniHsin2group}
$\A = 2\pi is \int_Z  \Tilde{B} \cup \Tilde{B}. $
See \eqref{'tHooft anomaly CS} and \eqref{Z_2 anomaly in U(1)_2m}.}
\begin{equation}
\label{Z_2 anomaly in U(1)_2m}
\A = \int_Z \frac{i2m}{4\pi} B \wedge B =
\frac{im\pi}{2} \int_Z \Tilde{B} \cup \Tilde{B} =
\frac{im\pi}{2} n, \qquad n=0,1.
\end{equation}
We see that, when $k=2m$, the $\Z_{2m}^{(1)}$ symmetry contains a
$\Z_2^{(1)}$ symmetry which
\begin{itemize}
\item is anomaly free when $m=0$ (mod 4). It is indeed generated by a
  bosonic line \eqref{spin middle line};
\item has a fermionic anomaly \eqref{Z_2^(d-2) anomaly} when $m=2$
  (mod 4). It is indeed generated by a fermionic line \eqref{spin
    middle line};
\item is generally anomalous for $m$ odd. In this case the
  anomaly is somewhat more severe and is associated to anyons
  \eqref{spin middle line}.
\end{itemize}
The second case is relevant for this work: $\U(1)_{4l}$ CS
theory, with $l$ odd, which is a bosonic theory, has a $\Z_2^{(1)}$
symmetry with the correct 't Hooft anomaly \eqref{Z_2^(d-2) anomaly}
to be the bosonic dual \eqref{fermionization} of a fermionic theory.

It is easy to see that gauging $\Z_2^{(1)} \subseteq \Z_{4l}^{(1)}$
yields $\U(1)_{4l} \ra \U(1)_l$. The most straightforward way is to
consider the four dimensional action \eqref{CS gauged in 4d with
  anomaly}, with level $4l$. Here, if $B$ is a $\Z_2$ gauge field,
$f-B$ is not a curvature for a $\U(1)$ connection, but $2f-2B$
is. Therefore, we can change variables to
$\Tilde{f} = \d \Tilde{a} = 2f -2B$ and rewrite \eqref{CS gauged in 4d
  with anomaly} as
\begin{equation}
  S= \int_Y \frac{i4l}{4\pi} (f-B) \wedge (f-B) =
  \int_Y \frac{il}{4\pi} \Tilde{f}\wedge \Tilde{f} =
  \int_X \frac{il}{4\pi} \Tilde{a}\d \Tilde{a}.
\end{equation}
Notice that this works for both $l$ even and $l$ odd, but in the
former case everything can be done also in the three dimensional
theory (there is no anomaly), while in the latter case the extension
to four dimensions is necessary (there is an anomaly). Indeed, after
gauging, we pass from a bosonic TQFT $\U(1)_{4l}$ to $\U(1)_l$: this
is bosonic for $l$ even, no issues arise, while is fermionic for $l$
odd. In this latter case a spin structure is required and this is
exactly the role of the anomaly $\A$ in \eqref{Z_2 anomaly in U(1)_2m},
accordingly to the general discussion of bosonization in Section
\ref{bosonizationSEC}. In other words: when $l$ is even, we start with
a gauge invariant bosonic theory $\U(1)_{4l}$ and, after gauging
$\Z_2^{(1)}$, we still get a well-defined theory $\U(1)_l$; when $l$
is odd, after gauging $\Z_2^{(1)}$, we obtain a theory which is
well-defined and gauge invariant only when a spin structure is
specified.\footnote{This can be done also with the three dimensional
  action: even if it is not precise, it yields the correct answer
  \cite{seibergwitten2016gappedTI}. Consider $\U(1)_{2m}$ CS and gauge
  $\Z_2^{(1)}$. Gauging $\Z_2^{(1)}$ does not change the gauge group,
  since $\U(1)/\Z_2 \cong U(1)$. However, $a$ is not a well-defined
  $\U(1)$ connection after this procedure, while $\Tilde{a} = 2a$
  is. Thus, under $\Z_2^{(1)}$ gauging, $2m \to m/2$.
  
When $k=4l$, thus $m =2l$, the final theory is $\U(1)_l$. If $l$ odd,
it requires a spin structure since the original $\Z_2^{(1)}$ symmetry
is anomalous. When $m$ is odd, the final theory is not
well-defined. This is a consequence of the anyonic nature of the
anomaly of $\Z_2^{(1)}$ which cannot be cured by simply introducing a
spin structure.}

We have just shown that $\U(1)_k$ CS is the fermionic dual of
$\U(1)_{4k}$, obtained by gauging the $\Z_2^{(1)}$ symmetry according
to the fermionization procedure \eqref{fermionization}. The inverse
map, namely bosonization \eqref{invertible bosonization}, is the
$\Z_2^f$ gauging: this allows to pass from $\U(1)_k$ to
$\U(1)_{4k}$. Let us check this. As said in Section \ref{U(1) CS is
  spin}, $\U(1)_k$ has a $\Z_2^{(0)}$ symmetry generated by the
reduction mod 2 of $\d a /2\pi$ which we identified with $\Z_2^f$. We
couple this symmetry to its $\Z_2$ background gauge field $s$ as
\begin{equation}
    S_{\U(1)_k}[s] = \int_X \frac{ik}{4\pi} a \d a + \frac{i}{2\pi} s\d a.
\end{equation}
The last term is to be intended as a mod 2 reduction of
$ks \d a /2\pi$ (which is indeed the coupling \eqref{CS depends on
  spin}). When $k$ is even, it is always trivial (coherent with the
fact that $\U(1)_{2n}$ is bosonic). The above action is thus relevant
for $k$ odd. To gauge this symmetry, we promote $s$ to a dynamical
gauge field, with a Lagrange multiplier $c$ to
ensure that we sum only on $\Z_2$ gauge field configurations, i.e.
\begin{equation}
  S_{\U(1)_k/\Z_2^f} = \int_X \frac{ik}{4\pi} a \d a +
  \frac{i}{2\pi} s\d a + \frac{2i}{2\pi} s \d c.
\end{equation}
Integrating out $s$ yields $a = -2c$ and so
\begin{equation}
    S_{\U(1)_k/\Z_2^f} = \int_X \frac{i4k}{4\pi} c \d c = S_{\U(1)_{4k}},
\end{equation}
as expected.

\newpage
\bibliographystyle{ieeetr}
\bibliography{bibliography.bib}

\end{document}